\documentclass[10pt]{article}

\usepackage{amssymb,amsfonts,amsmath,bm}
\usepackage{graphicx}
\usepackage[usenames]{color}

\usepackage{xr}

\usepackage{array}

\usepackage{color}

\usepackage{cite}

\usepackage{capt-of}
\usepackage{rotating}

\catcode`\_=\active
\def_#1{\ensuremath{\sb{\text{#1}}}}

\usepackage{setspace} 
\usepackage[labelfont=bf,labelsep=period,justification=raggedright]{caption}

\makeatletter
\renewcommand{\@biblabel}[1]{\quad#1.}
\makeatother

\date{}

\begin{document}

\begin{flushleft}
{\Large
\textbf{Separability of drag and thrust in undulatory animals and machines}
}
\\
Rahul Bale,$^{1,+}$ Anup A. Shirgaonkar,$^{1,+}$ Izaak D. Neveln,$^{2}$  Amneet Pal Singh Bhalla,$^{1}$ \\
Malcolm A. MacIver,$^{1,2,3, \ast}$  and Neelesh A. Patankar$^{1,\ast}$\\

\bf{1} Department of Mechanical Engineering, Northwestern University, 
\\
\bf{2} Department of Biomedical Engineering, Northwestern University, 
\\
\bf{3} Department of Neurobiology, Northwestern University, \\
$\;$ $\;$ 2145 Sheridan Road, Evanston, IL 60208, USA

$+$ Authors made equal contributions

$\ast$ Authors for correspondence (n-patankar@northwestern.edu, maciver@northwestern.edu)
\end{flushleft}

\graphicspath{{../figs/ai-tweaked/}}

\section*{Abstract}
For nearly a century, researchers have tried to understand the
  swimming of aquatic animals in terms of a balance between the
  forward thrust from swimming movements and drag on the body.
  Prior approaches have failed to provide a separation of these two forces for undulatory swimmers
  such as lamprey and eels, where most parts of the body are simultaneously
  generating drag and thrust. 
We nonetheless show that this separation is possible,
  and delineate its fundamental basis in undulatory swimmers.
  Our approach unifies a vast diversity of undulatory aquatic animals (anguilliform, sub-carangiform, gymnotiform, balistiform, rajiform) and provides design principles for highly agile bioÐinspired underwater vehicles.
  This approach has practical utility within biology as well as engineering. It is a predictive tool for use in understanding the role of the mechanics of movement in the evolutionary emergence of morphological features relating to locomotion. For example, we demonstrate that the drag-thrust separation framework helps to predict the observed height of the ribbon fin of electric knifefish, a diverse group of neotropical fishes which are an important model system in sensory neurobiology. 
  We also show how drag-thrust separation leads to models that can predict the swimming velocity of an organism or a robotic vehicle.  

\section{Introduction}
The hydrodynamics of aquatic locomotion has importance across multiple domains,
from basic biology to the engineering of highly maneuverable underwater vehicles.
Within biology, beyond
the interest in aquatic locomotion, 
there is extensive use of several aquatic model systems within neuroscience
such as lamprey for research into spinal cord function \cite{Gril03a}, 
zebrafish for developmental 
neuroscience \cite{Mcle08a}, and weakly electric fish for the neurobiology of
sensory processing (reviews: \cite{Turn99a,Krah13a}). Within engineering, the 
maneuverability and efficiency of fish is inspiring new styles of
propulsion and maneuvering in underwater vehicles \cite{MacI04a,Neve13a,Colg04a}. The
implementation of engineered solutions will depend on the
resolution of open issues in hydrodynamics of aquatic locomotion. 
Finally, an understanding of the hydrodynamics of aquatic
locomotion is critical for insight
into the evolution of fish \cite{Webb84a} and
their land-based descendants.

While the hydrodynamics of swimming organisms have
been studied actively for almost a century, there are key questions 
that remain unresolved. This work focuses on one
such unresolved issue that pertains to the mechanisms of drag
and thrust generation. The decomposition of the total force
on a swimming organism into drag and thrust is 
desirable because it can fundamentally reveal how an organism 
produces forward push to balance the resistance to motion
from the surrounding fluid. It can also lead to simple quantitative
models to predict swimming velocity of organisms 
or artificial underwater vehicles based on their kinematics.

\begin{figure}
\centerline{\includegraphics[width=3.0in]{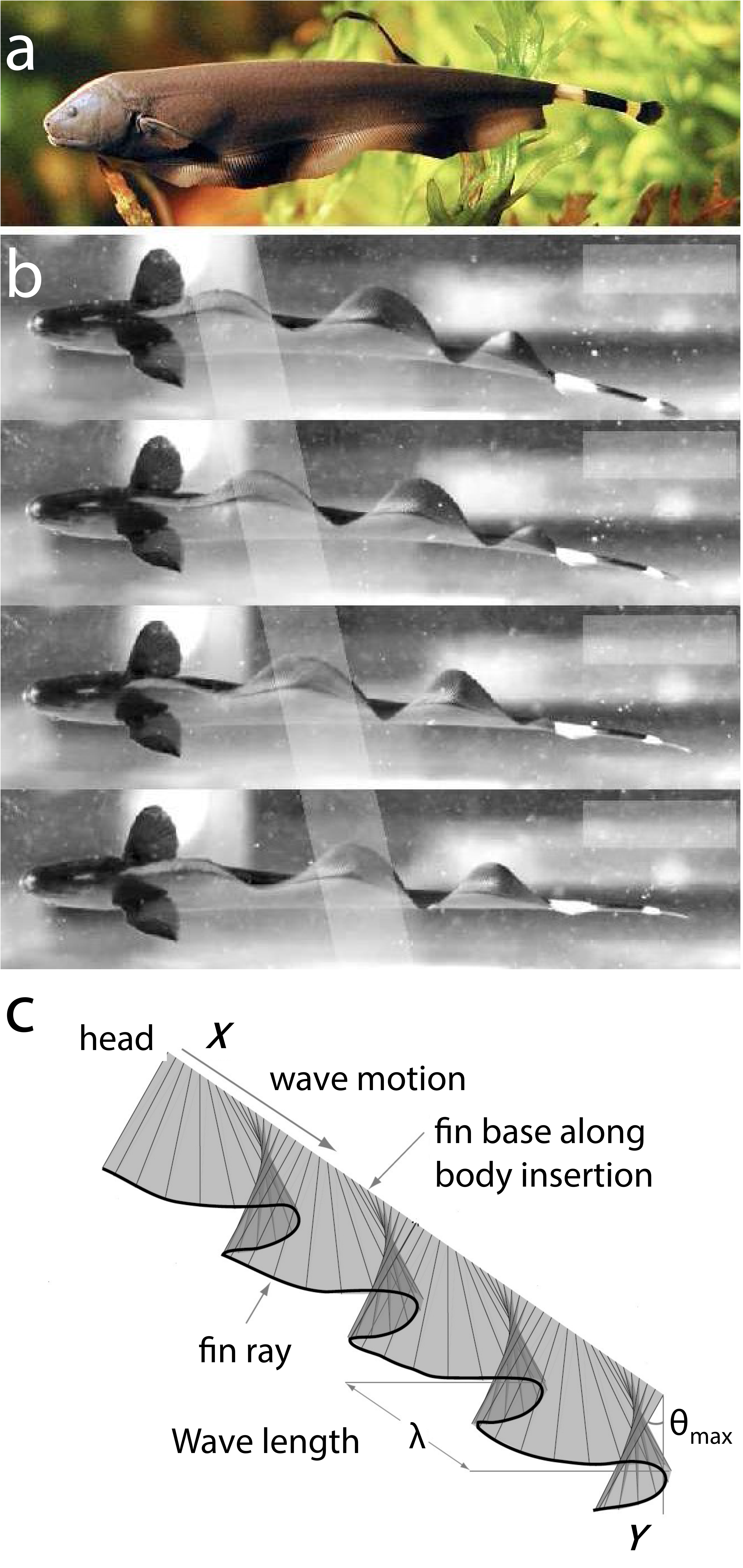}}
\caption{a) A median fin (ribbon fin) undulatory swimmer (gymnotiform swimmer): \textit{Apteronotus albifrons}, the black ghost knifefish of South America (photograph courtesy of
Per Erik Sviland). b) A backward traveling wave on the ribbon fin. c) Geometric configuration of the ribbon fin without the
body for computations. Figure adapted from Fig.~1 of \cite{Ruiz13a}.} 
\label{fig:gymbali}
\end{figure}

If we
consider a boat with a propeller, the decomposition of thrust and drag is straightforward since all the thrust is coming from the propeller, and
most of the drag is coming from the hull. However, in the
case of anguilliform swimmers such as eels, where the entire
body undulates, there are no distinct portions of the body that alone produce
thrust or cause drag. In other groups of fishes 
there are undulatory elongated fins along the
ventral midline (Fig.~\ref{fig:gymbali}a, knifefish such as those of the Gymnotiformes 
and Notopteridae), 
dorsal midline (\emph{Gymarchus nilotics}, the oarfish \emph{Regalecus glesne}), 
along ventral and dorsal midlines
(triggerfish of the Balastidae), 
and along the lateral margins of the body (certain rays and skates of the 
Batoidea, cuttlefish).
For these animals, undulating fins may be regarded as the primary thrust
generators and the relatively straight body may be regarded as the primary source of drag.
However, this apparent decomposition of drag and thrust regions 
should not be considered to imply that an undulatory fin 
itself has no drag. This is clearly not the case because a 
hypothetical undulatory ribbon fin that is not attached to a body
will undergo steady swimming. The drag of the ribbon fin and the thrust
it generates will be in balance in that case.

Arguably the most widely cited analysis of drag-thrust
decomposition is due to Lighthill \cite{Ligh75a}. He considered the force
generated from undulatory motion by elongated animals such
as eels \cite{Ligh71a}, and later, with Blake, considered forces on 
ribbon fins of balistiform and gymnotiform swimmers \cite{Ligh90a}. The
analysis was based on a ``reactive'' theory of propulsion that
was proposed for high Reynolds number swimming. Lighthill
considered a decomposition of the force on the body into resistive 
and reactive components that lead to drag and thrust,
respectively. He defined the force on a section of the body as
resistive if it depends, linearly or non-linearly, on the instantaneous 
velocity of that section relative to the surrounding fluid.
Reactive forces were defined as those due to inertia of the 
surrounding fluid (the ``added mass'' effect), 
proportional to the rate of change of the relative velocity
between the fluid and surface of the swimming body. Lighthill
then provided expressions for the reactive thrust force using a
simplified potential flow theory \cite{Ligh71a,Ligh90a}. No model for drag was
developed.

To address this long standing issue, we define three fundamental
principles for the decomposition of forces arising from undulatory
swimming into drag and thrust.
First, (D1) the body movement (kinematics) creating drag needs to
be separated from the body movement creating thrust, such that the
sum of these two movements results in the originally observed
swimming motion of the animal. Second, (D2) these decomposed movements
should be such that the surface of the body will move in a continuous
fashion so that the kinematics can be realized in experiments or
simulations. Third, (D3) the sum of the force due to the drag--inducing
movement with the force due to the thrust--inducing movement needs
to equal the force estimated from the original (undecomposed) movement of the fish. 
Requirement (D2) will enable three independent experiments to estimate forces 
and develop predictive models for swimming: i) the force based on the undecomposed motion; 
ii) the force resulting purely from the drag kinematics, defined as drag; 
iii) the force resulting purely from the thrust kinematics, defined as thrust. 

In this work we propose a new way to decompose drag and thrust that satisfies conditions D1 to D3. 
Using an idealized elongated anal fin (hereafter ribbon fin)
of weakly electric fish (Fig.~\ref{fig:gymbali}a) but with no body as a model 
system, we present results from simulations and experiments supporting our approach. 
The scope of applicability of the decomposition 
will be further demonstrated through simulations of the eel \emph{Anguilla rostrata}, 
the larval zebrafish \emph{Danio rerio}, 
the black ghost knifefish 
\emph{Apteronotus albifrons}, 
and the mackerel \emph{Scomber scombrus}.  
These examples show where the decomposition is valid, and where it becomes invalid.
For a case where it is valid---swimming with elongated fins such as in the 
knifefish---we go on to show that drag thrust decomposition can be used 
to predict an important morphological feature: the height of the fin that maximizes cost of transport.
Our predictions agree well with the measured height of the fin in a sample of 13 species in a representative family of knifefishes, the Apteronotidae.

\section{Results}

\subsection{Kinematic decomposition}
\label{sec:method:kin-decomp}
To demonstrate drag--thrust decomposition, we consider the same
model problem considered by Lighthill and Blake \cite{Ligh90a}
and analyze the forces on the elongated median fin (hereafter
``ribbon fin") of a gymnotiform swimmer (Fig.~\ref{fig:gymbali}a).
We numerically simulate a translating ribbon fin with a traveling
wave (Fig.~\ref{fig:gymbali}b) along it. The traveling wave on the 
ribbon fin was described by the angular position of any point on the ribbon fin  
$\theta(x,t) = \theta_{max} \sin2\pi(\frac{x}{\lambda}- f t)$, 
where $\theta_{max}$ is the maximum angle of excursion, $f$ is the 
frequency, and $\lambda$ is the wavelength of undulations.
In the simulations there is no attached body.
The fin morphology is shown in Fig.~\ref{fig:gymbali}c.
The force on the ribbon fin from the
fluid is numerically computed for different values of the translational
velocity $U$ of the fin and the traveling wave velocity $U_{w}$ (given by $f\lambda$).
The wave motion is caused by the lateral oscillatory velocity field $V_{w}$ on the fin surface. $V_{w}$, and by
consequence $U_{w}$, were varied by changing the frequency $f$ of
the traveling wave. 
The fin had two waves along its length \cite{Blak83a, Ruiz13a}. 

\begin{figure}
\centerline{\includegraphics[width=.99\textwidth]{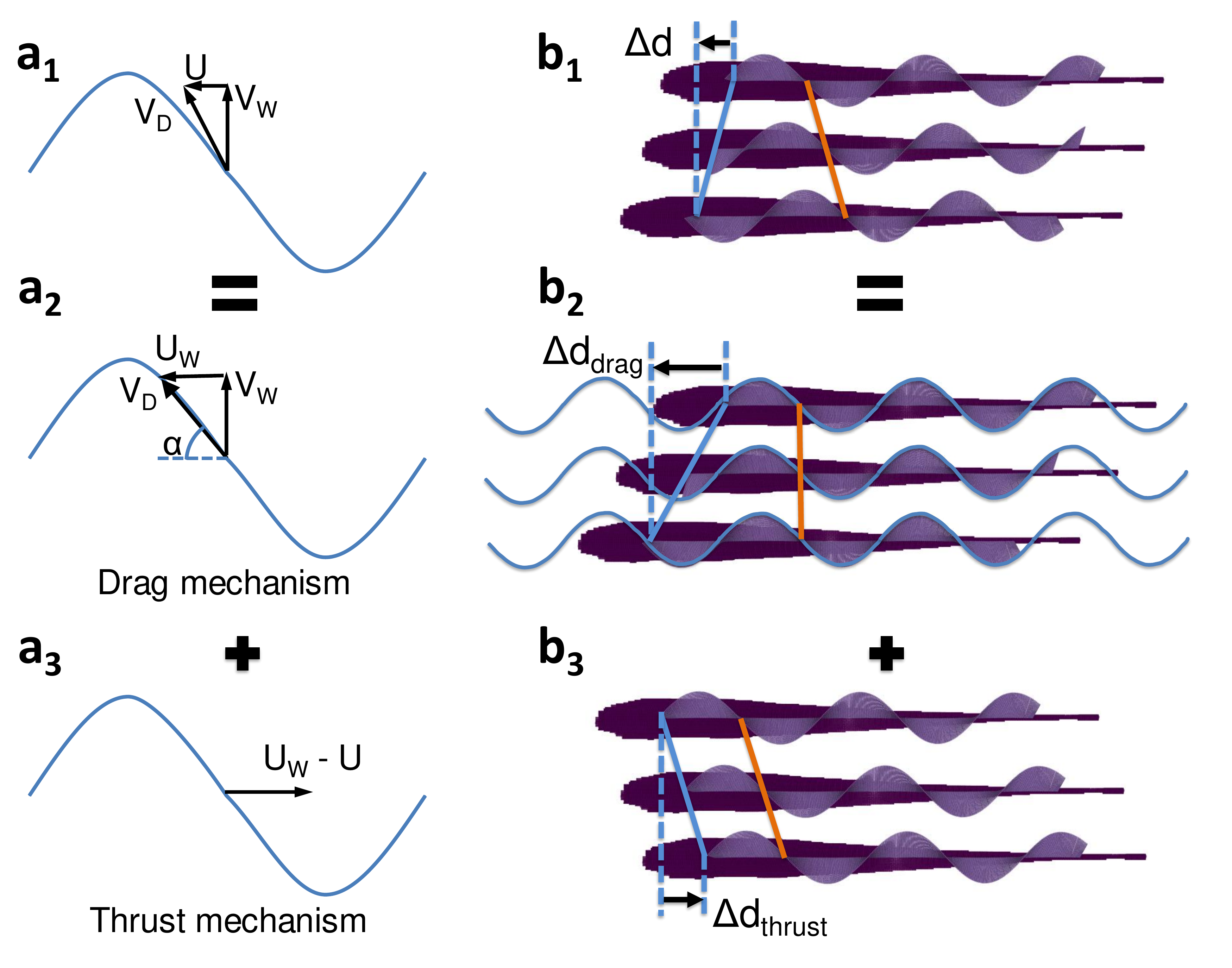}}
\caption{a) The proposed kinematic decomposition into drag and thrust
producing mechanisms. The front end is on the left. b) The kinematic 
decomposition applied to a modeled black ghost knifefish. Each row, from 
top to bottom in each of the undecomposed, drag ,and thrust mechanism, 
is at increasing instants of time.}
\label{fig:kd}
\end{figure}


To understand our proposed kinematic decomposition, consider
a generic waveform of constant amplitude that has a lateral
oscillatory velocity $V_{w}$ and a corresponding traveling wave
velocity $U_{w}$ (Fig.~\ref{fig:kd}-a1 \& a2). Let $U$ be the forward
velocity of the fin as a whole. This is the undecomposed kinematics which is 
decomposed into two parts (Fig.~\ref{fig:kd}-a2 \& a3). 
The first decomposed motion, which we term the drag--causing
\emph{perfect slithering} motion, occurs when the forward velocity
of the fin is equal to the backward velocity of the traveling wave ($U_{w}$).
As a result, the fin appears to move along a stationary wave--shaped
track (Fig.~\ref{fig:kd}-b2). Each point on the fin has a velocity
that is tangential to its surface ($V_D$ in Fig.~\ref{fig:kd}-a2). In
this case the fluid is dragged forward by the tangential velocity
along the fin surface which results in a backward (drag) force on
the fin from the fluid. The second decomposed motion, which we term
the thrust generating \emph{frozen fin} motion, occurs when the fin is
frozen in its undulatory shape, and that frozen shape drifts backward
with a velocity equal to the velocity of the traveling wave minus
the translational velocity of the fin $(U_{w}-U)$, when $U_w$ is
greater than $U$. This motion pushes the fluid backward which in
turn produces a forward (thrust) force on the fin (Fig.~\ref{fig:kd}-a3 \& b3).

In the idealization of the traveling wave as a sinusoid of constant
amplitude, superposition of the kinematics of the perfect slithering
and frozen fin motions results in the undecomposed kinematics of
the fin surface. This fulfills both condition D1 (separated body movements add up to give original body movements) and
condition D2 (the separated body movements are without discontinuities 
and physically realizable). This kinematic decomposition would be
exact in case of an infinitely long fin. However, this is not the case for
the finite
fin length that we consider. For example, in the frozen fin case,
the wave can be considered frozen in different phases. Despite this,
our computational fluid simulations show that the thrust force does
not depend strongly on the phase as long as there is more than one
full wave on the fin, as is the case here.  We next show that the
kinematic decomposition also fulfills the requirement that the force
of drag and the force of thrust sum to the undecomposed force from
the fin (D3).

\subsection{Dynamic decomposition}
\begin{figure}
\centerline{\includegraphics[width=.89\textwidth]{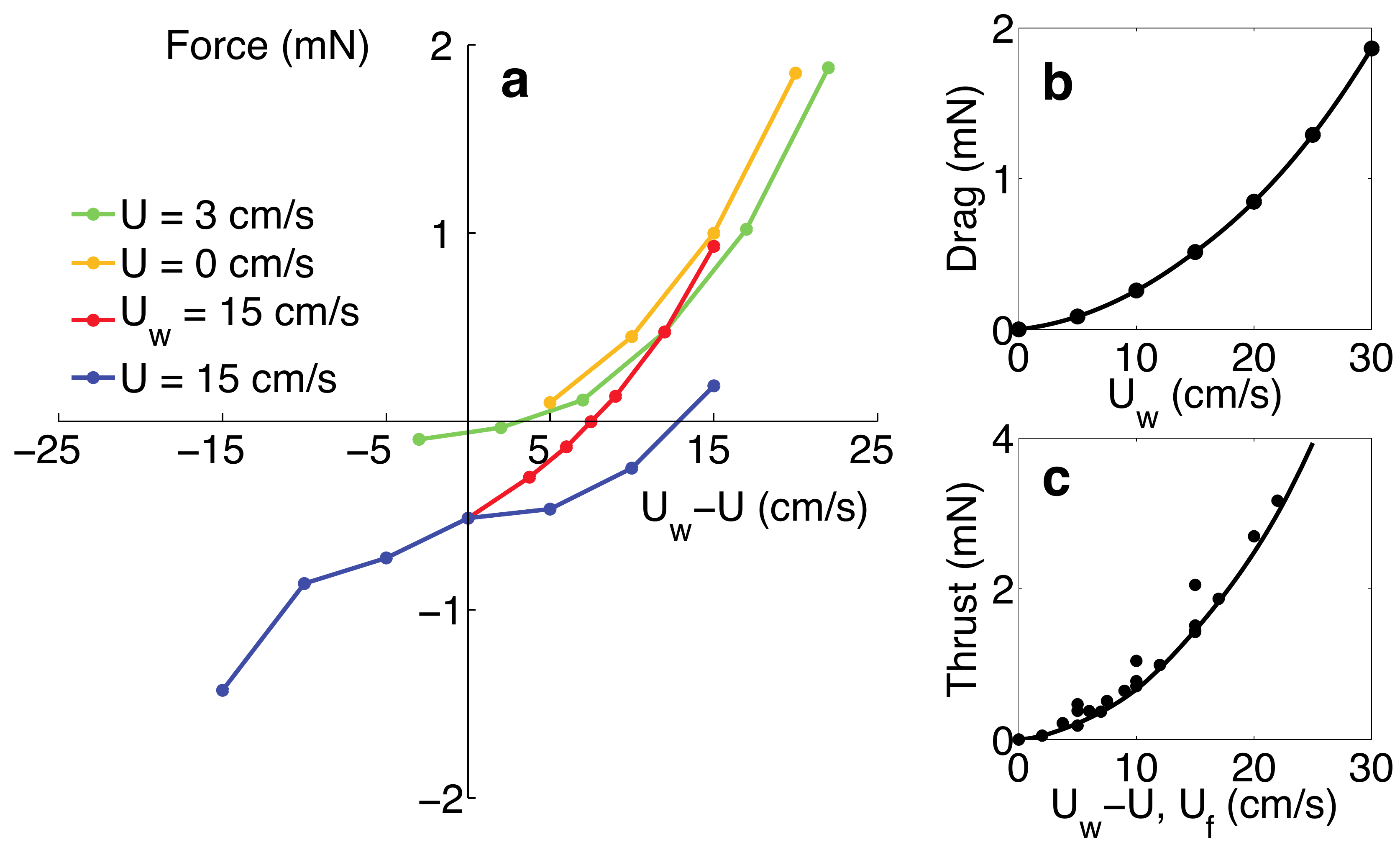}}
\caption{a) The computed total force on the ribbon fin vs.
$U_{\text{w}}-U$ for different parameters. The legend identifies different simulation sets. For example, 
the set with $U_{\text{w}}=15$~cm/s was the one where $U_{\text{w}}$ was fixed and the value of 
$U$ was changed. b) Drag , i.e., the force on a ribbon fin during
perfect slithering motion. c) Thrust $T$ computed as a function of $U_{\text{w}}-U$
for each data point in (a) by assuming the
kinematic decomposition (Fig.~\ref{fig:kd} and Eqn.
\ref{eqn:decomp}). These data are shown by solid dots. Separate
frozen fin simulations were conducted as a function of $U_{f}$,
shown by the dashed line.  The dots cluster along this line
giving evidence for the successful decomposition of drag and
thrust.} 
\label{fig:forces}
\end{figure}

Dynamic decomposition here implies the decomposition of forces on the swimming body. 
Without loss of generality we consider a traveling wave that is
moving backward, as in Fig.~\ref{fig:gymbali}b.
If the decomposition into drag and thrust is valid,
then the total force $F$ on the fin should satisfy the following
equation
\begin{equation}
F[\;U,\; U_{w}\;] = \textsf{sgn}[\;U_{w}-U\;]\;\;T[\;U_{w}-U\;] - D[\;U_{w}\;],
\label{eqn:decomp}
\end{equation}
where square brackets indicate ``function-of'', $T$ is thrust, $D$ is drag, and
$\textsf{sgn}[\cdot]$ gives the sign of the argument. The data for
$F[U, U_{w}]$ are plotted in Fig.~\ref{fig:forces}a. 

We performed a separate set of simulations for the perfect slithering motion
($U=U_{w}$) for different values of $U_{w}$. The force on the
ribbon-fin in this case is $D[U_{w}]$, which is plotted in Fig.~\ref{fig:forces}b.

Using Eqn. \ref{eqn:decomp} and the results in Fig.~\ref{fig:forces}b 
 we calculate the thrust $T[U_{w}-U]$ for each data point in
Fig.~\ref{fig:forces}a. If the decomposition is valid then the
resulting data should be a well defined function of $U_{w}-U$. 
This is found to be so in Fig.~\ref{fig:forces}c. Additionally, 
the results for $T[U_{w}-U]$, obtained above, should also match 
results from another set of simulations for the frozen fin case. 
To check this we performed frozen fin simulations for different 
values of the fin translational velocity $U_{f}$. There is no 
traveling wave in this case. The force on the
ribbon fin in this case is $T[U_{f}]$ which should be the same
function as $T[U_{w}-U]$ with $U_{f}$ replaced by $(U_{w}-U)$. The
solid line of Fig.~\ref{fig:forces}c confirms this expectation.

Thus, we have two new results: first, an approach to
separate the mechanisms of drag and thrust, and second we
obtain a correlation not only for the thrust (Fig.~\ref{fig:forces}c) but
also for the drag (Fig.~\ref{fig:forces}b) on an undulatory propulsor.

\begin{figure}
\centerline{\includegraphics[width=.89\textwidth]{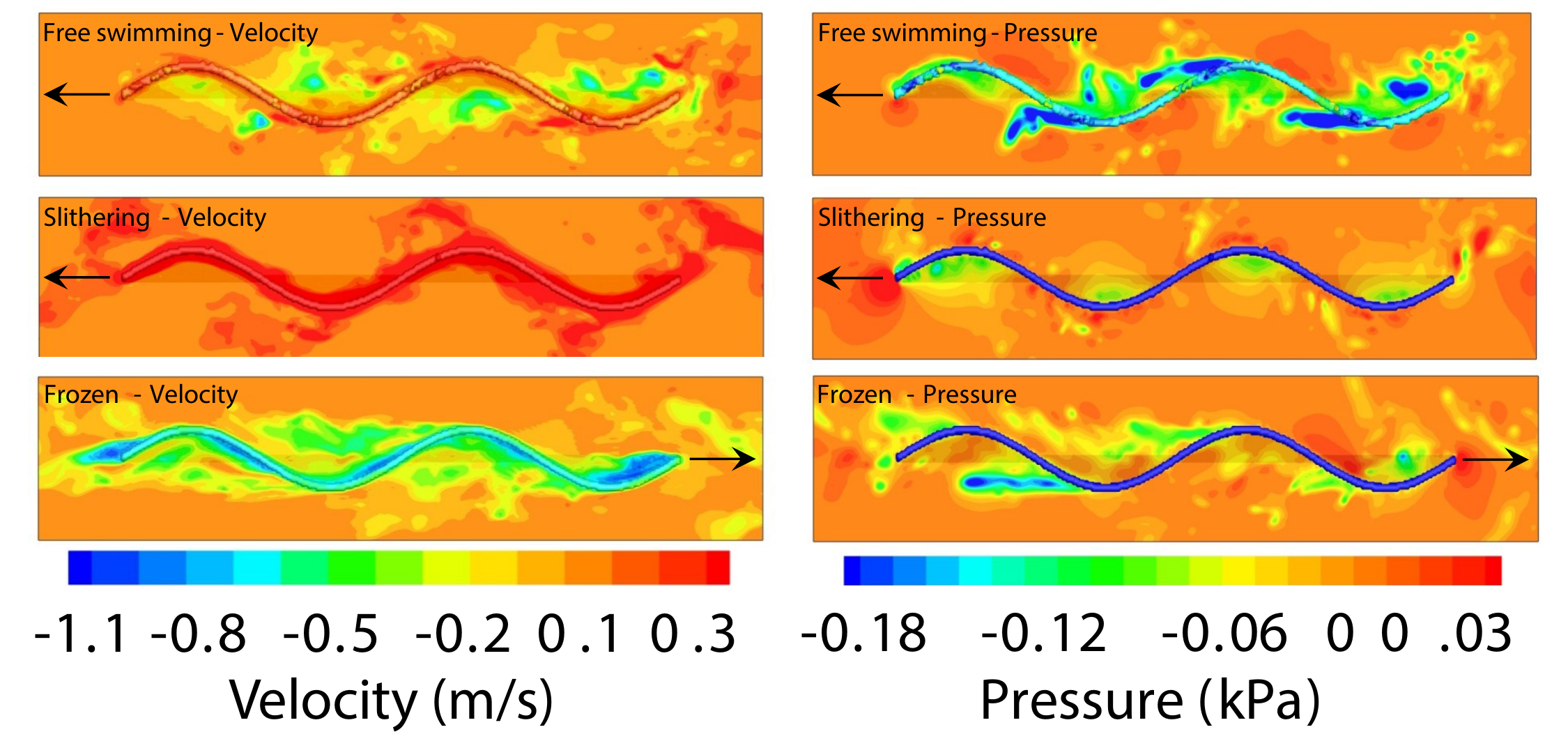}}
\caption{The axial velocity (left) and pressure (right) fields in
a cross-sectional plane at the bottom edge of the ribbon fin for
three cases. Top: Normal case with $U=3$~cm/s and $U_{\text{w}}=15$~cm/s.
Middle: Perfect slithering motion with $U=15$~cm/s and
$U_{\text{w}}=15$~cm/s. Bottom: Frozen fin motion with a backward (i.e.
to the right) velocity of $12$~cm/s. The legend for axial velocity
(left) show magnitudes that are scaled by $U_{\text{w}} = 15$~cm/s. In
the contour plot velocity to the left (i.e. forward direction) is
positive and to the right (backward) is negative. The legend for
pressure (right) show magnitudes scaled by $\rho U_{\text{w}}^2$ where
$U_{\text{w}} = 15$~cm/s.} \label{fig:velpress}
\end{figure}

\subsection{The spatial segregation of drag- and thrust-related flows}
Consider a ribbon fin moving
with $U=3$~cm/s and $U_{w}=15$~cm/s. 
Through simulation, the total forward force is found to be $0.46$~mN. 
The drag causing perfect slithering mode has $U=15$~cm/s and $U_{w}=15$~cm/s. The thrust generating frozen mode has no wave velocity but has a backward
velocity of $U_{w}-U = 12$~cm/s. 
Separate simulations were conducted for the drag and thrust causing modes. 
The calculated thrust and drag forces were $0.92$~mN and $0.52$~mN, respectively.
The difference is $0.4$~mN which is close to $0.46$~mN computed
for the un-decomposed case, i.e., Eqn. \ref{eqn:decomp} is
approximately satisfied.
We plot the simulated axial velocity and pressure for these three cases in
Fig.~\ref{fig:velpress} for the same phase of the fin. 
The slithering drag mode has thin boundary layers
outside of which the velocity has low magnitude and does not
have strong spatial gradients. On the other hand, the 
frozen thrust mode has strongly separated regions behind the
troughs and crests of the wave along the fin. This velocity field and its
gradients are significant \emph{outside} the boundary layer region
of the slithering mode. Wherever the velocity due
to one mode is high, the velocity due to the other mode is
low. The coupling of the drag and thrust causing
modes, through the nonlinear inertia term ($(\bm{u}\cdot\nabla)\bm{u}$) in the Navier--Stokes equations (equation \ref{eqn:NS}), 
is weak. Furthermore, Fig.~\ref{fig:velpress} shows that the dominant pressure regions due to the two modes are also spatially segregated. 
The low pressure due to the
thrust-causing frozen fin mode is dominant behind the wave troughs
and crests, whereas the low pressure due to the slithering mode is
dominant away from the separation region in the concave part of
the wave shape.
This spatial segregation of the drag-- and thrust--related flows is the
fundamental basis of the success of the drag--thrust decomposition.

Finally, we note the contributions to the
force from pressure and viscous terms. Due to separation, the
pressure contribution to the thrust force dominates in the thrust
causing frozen fin mode. Of the total thrust force of $0.92$~mN,
the pressure contribution is $0.81$~mN and the remainder is due to the 
viscous contribution. In the drag causing
slithering mode the viscous contribution is $0.12$~mN out of the
total drag force of $0.52$~mN and the remainder is due to the pressure contribution. 
Thus, the pressure force dominates thrust
while the viscous contribution to drag is relatively large due to
thin boundary layers.

It has long been hypothesized that the drag of swimming fish is
higher due to the thinning of the boundary layers caused by
undulatory motion \cite{Ligh75a}. Fig.~\ref{fig:velpress} shows
that the boundary layer flow is a key feature of the drag
causing slithering mode. 
Separated flow
plays a role in the thrust mechanism. The separated flow regions
are suction zones where the fluid is sucked backward by
the undulating fin. This leads to the thrust force. This is consistent 
with our flow visualization data reported earlier \cite{Shir08a,Neve13b}.

\subsection{Generality of the drag--thrust decomposition}
\label{sec:appl}
The decomposition described above was applied to an idealized ribbon-fin with idealized kinematics at a Reynolds number around 10,000. 
This idealization may not be valid for swimming animals, thus we
now examine the applicability of the drag--thrust decomposition to
swimming animals with realistic body/fin geometries and measured kinematics.
We also examine the validity of the decomposition at moderately
high Reynolds numbers by applying the decomposition to a robotic
knifefish.

\subsubsection{Application to swimming animals}
Our kinematic decomposition of drag and thrust
assumed a constant amplitude wave (Fig.~\ref{fig:kd}). This
assumption is not strictly valid for swimming animals. For example, in the black ghost
knifefish the amplitude of oscillation of the ribbon-fin
tapers-off toward the two ends \cite{Ruiz13a}. 
Anguilliform and carangiform swimmers have an amplitude that 
increases with body length \cite{Vide84a,Tyte04a,Bora08a}. 
Additionally, the wave motion may
not be strictly sinusoidal. In the non-constant amplitude case, the kinematic split 
as proposed in this work will not be exact. However, if the amplitude changes
are not large then the additional error may not be significant.
The approach is expected to work in cases where cross-sections (width of the body in the lateral direction) of the body are
non-uniform, provided that the body
width does not change sharply along the body.
The proposed decomposition is not expected to work for
indefinitely high Reynolds numbers, but it does work at moderately high Reynolds numbers which will be demonstrated 
in the next subsection. Finally, as the height of the 
ribbon-fin and its amplitude of oscillation is reduced, the drag and
thrust producing flow fields may not remain as separate as the case shown in 
Fig.~\ref{fig:velpress}. Thus, as the undulatory propulsor
becomes slender the decomposition of drag and thrust may not be as
clear. Given these issues, we examine the applicability of the decomposition 
with simulations of the eel \emph{Anguilla rostrata}, 
the larval zebrafish \emph{Danio rerio}, 
the black ghost knifefish 
\emph{Apteronotus albifrons}, 
and the mackerel \emph{Scomber scombrus}.  
These examples show where the decomposition is valid, and where it becomes invalid.

\newcolumntype{M}[1]{>{\centering\arraybackslash}m{#1}}
\renewcommand{\arraystretch}{2}

\begin{table}
\centering

 \begin{tabular}{M{2.8cm}  M{2.8cm}   M{2.8cm} M{2.8cm}  M{2.8cm}}  \hline \hline
  
                  & \textbf{Knifefish} & \textbf{Eel} & \textbf{Zebrafish (larval)} & \textbf{Mackerel}\\ \hline \hline
                              
    \textbf{Amplitude (cm)}   &  \includegraphics[scale=0.11]{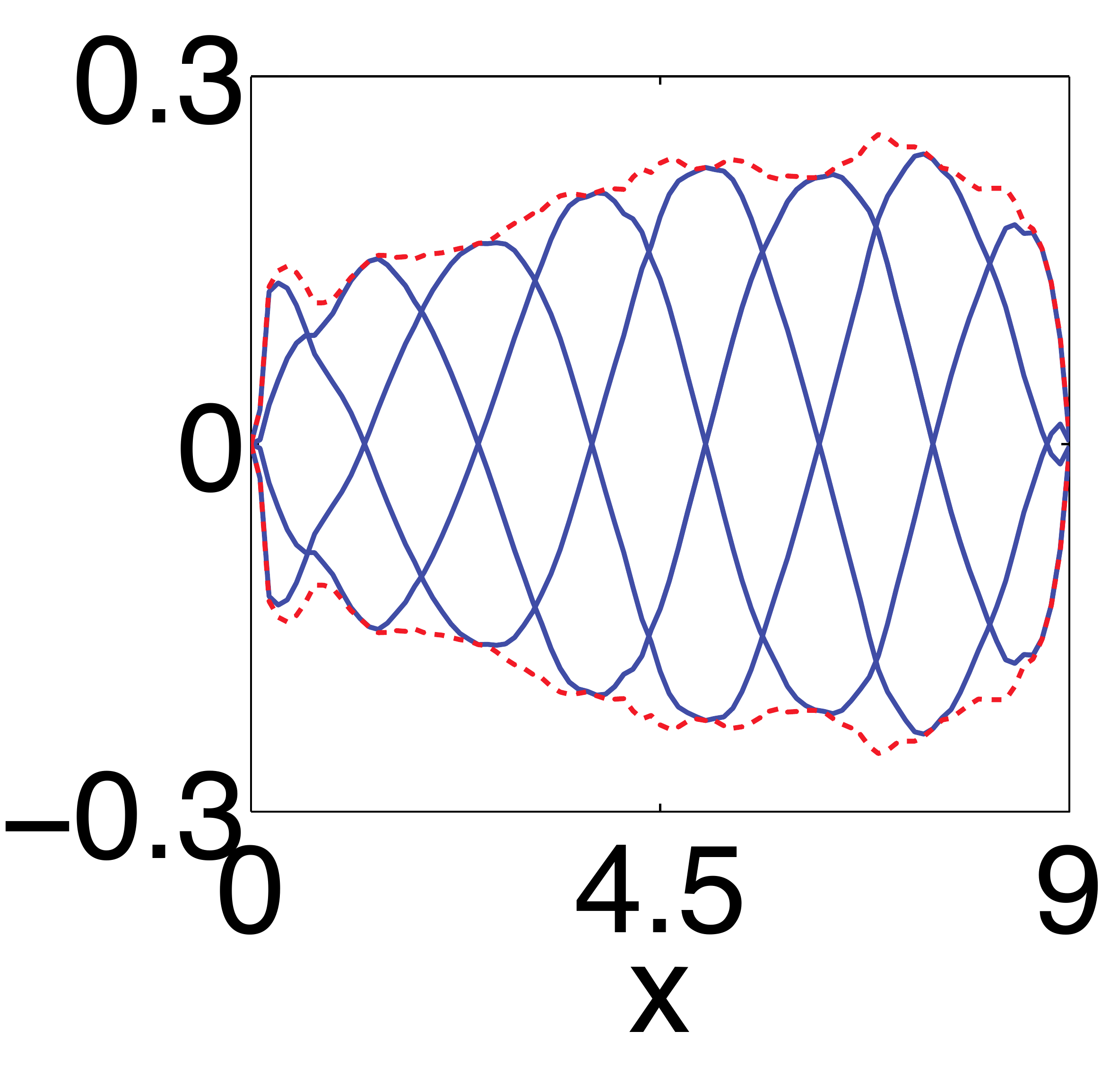}  
                              & \includegraphics[scale=0.11]{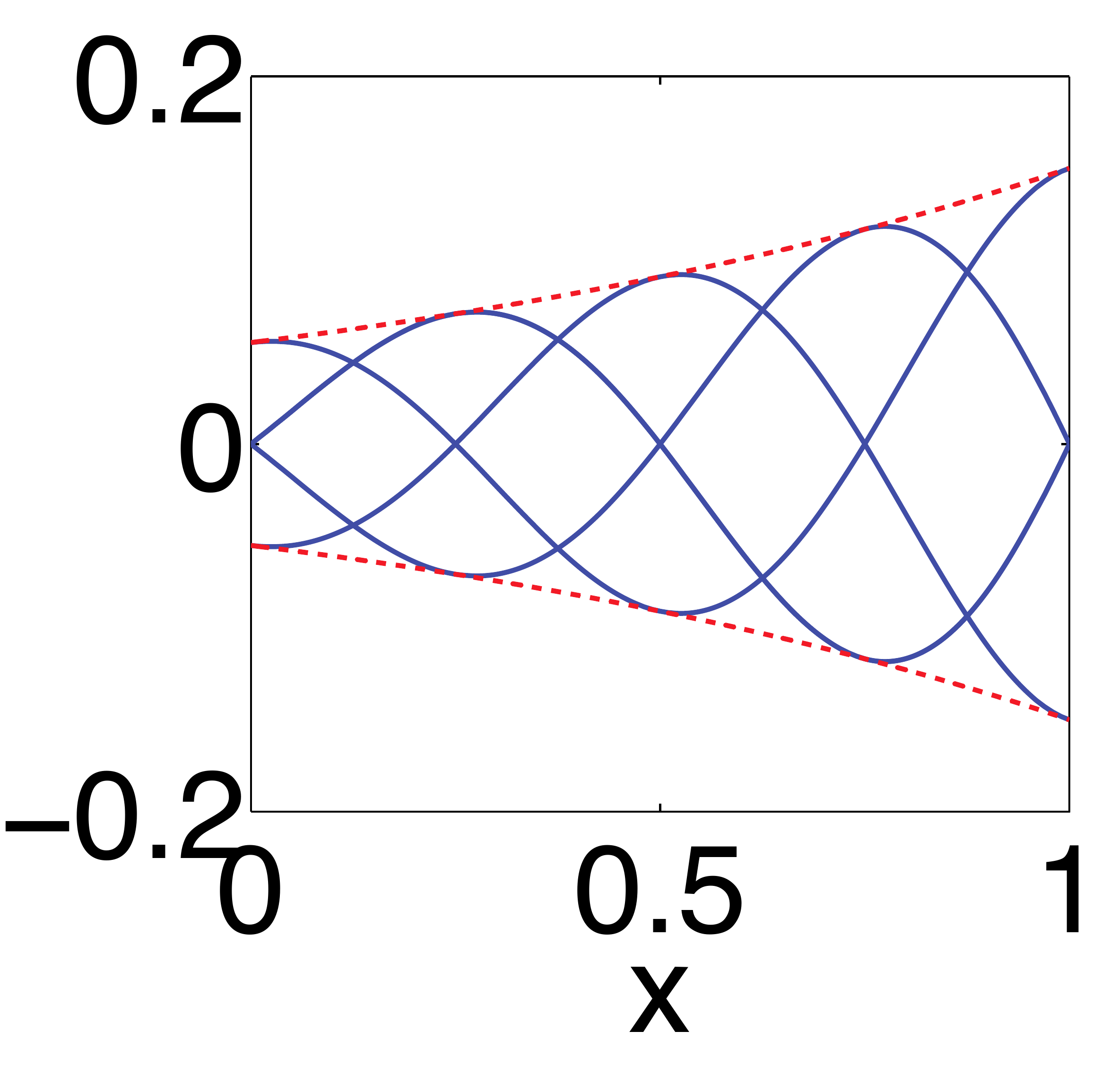} 
                              & \includegraphics[scale=0.11]{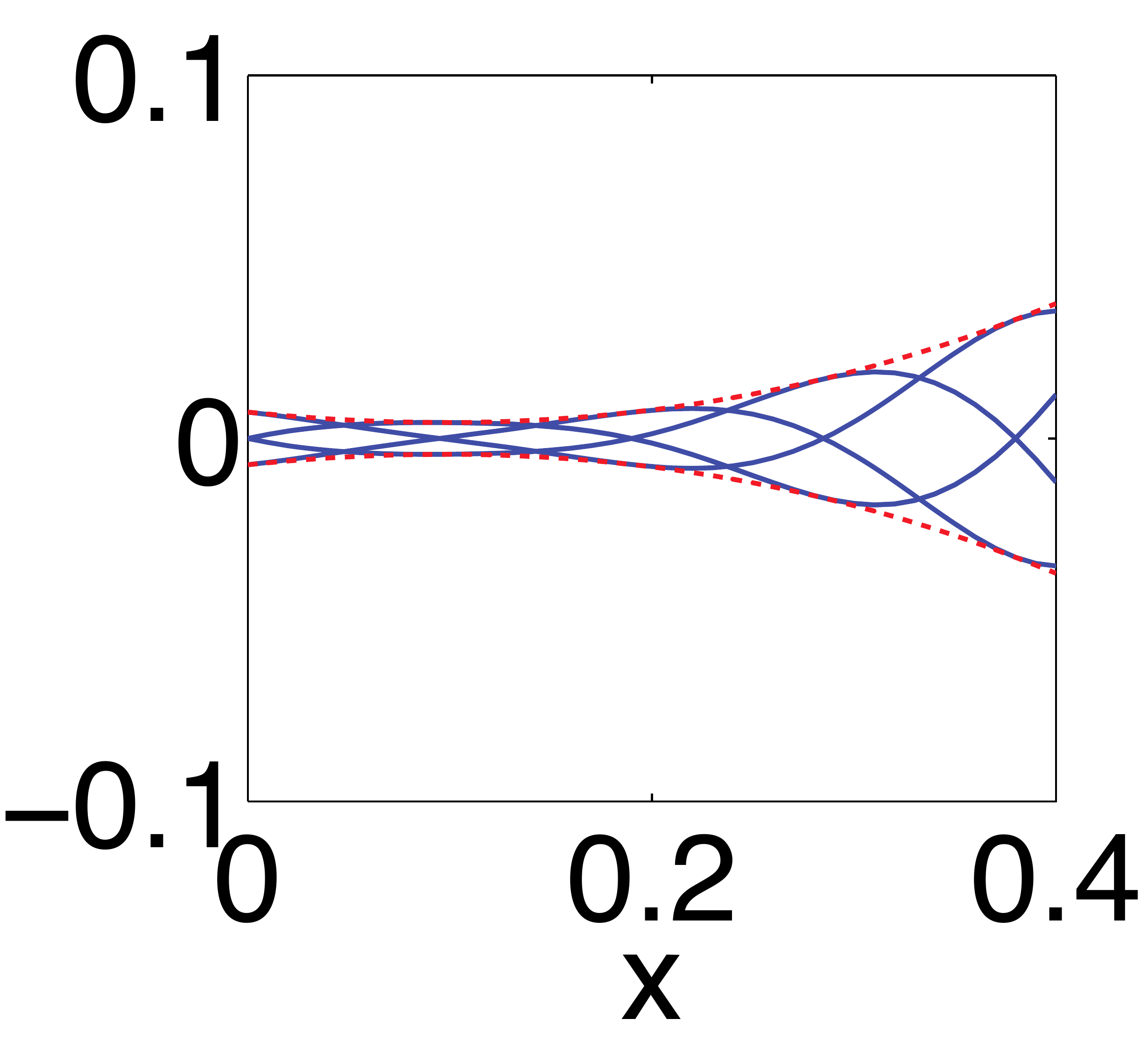} 
                              & \includegraphics[scale=0.11]{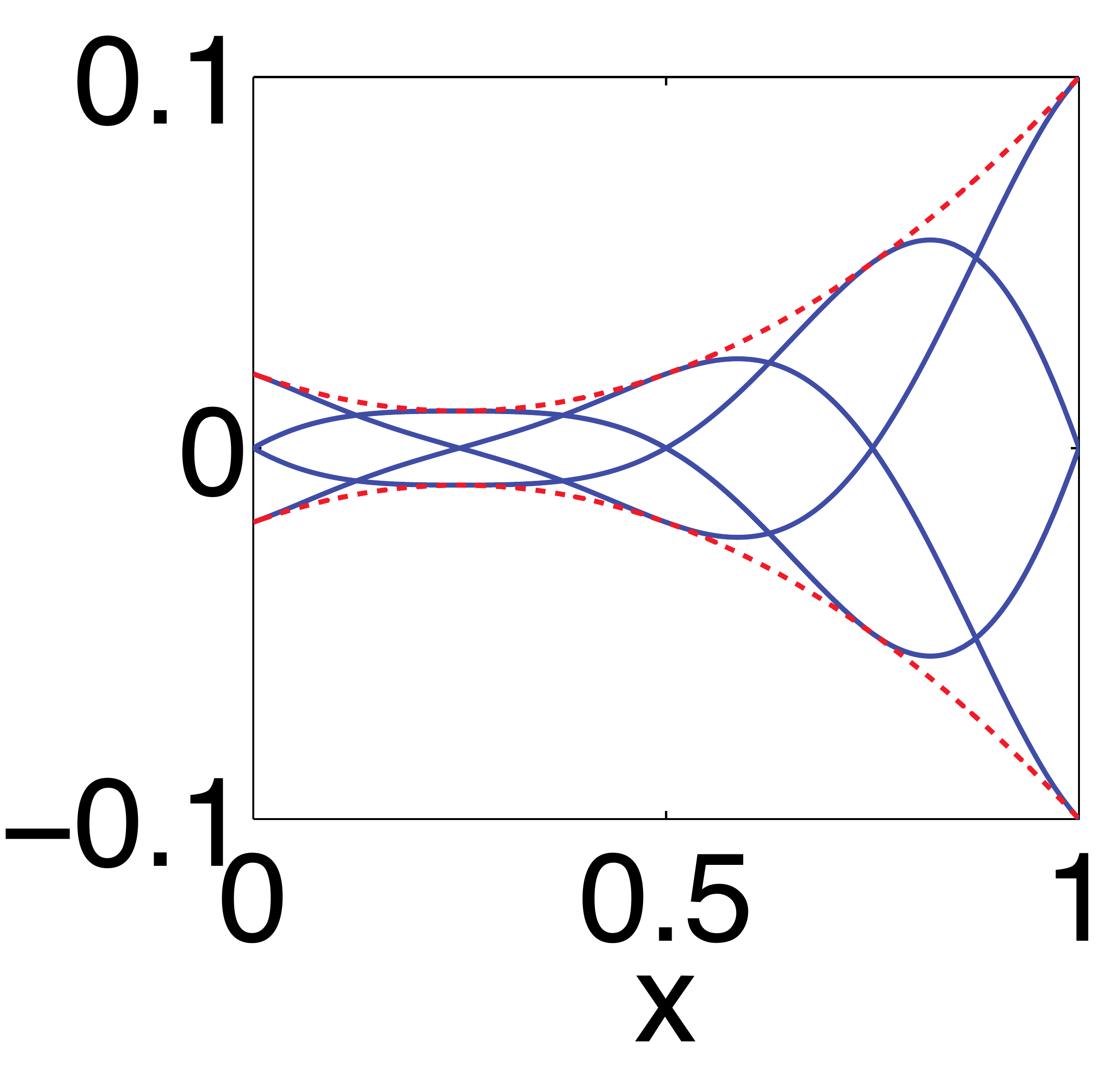} \\

    \textbf{Amplitude function $A(x)$} & Experimental~\cite{Ruiz13a}  & $0.15 e^{(x-1)}$\cite{Tyte04a} & Experimental &  $0.02-0.08x+0.16x^2$\cite{Vide84a, Bora08a}  \\
    \textbf{Thrust force $T$ (mN)}  &   1.43  &  6.8$\times 10^{-4}$  & 4.3$\times 10^{-3}$ & 11.6$\times 10^{-3}$\\ 
    \textbf{Drag force $D$ (mN)}   &   1.48  &  8.2$\times 10^{-4}$ & 4.1$\times 10^{-3}$  & 46.8$\times 10^{-3}$ \\ 
    \textbf{ Undecomposed force $F$ (mN) } & 0   & 0  & 0 & 21.4$\times 10^{-3}$ \\    
    \textbf{Net force $D-T$ (mN)}   &   0.05  &  1.4$\times 10^{-4}$ & $-$0.2$\times 10^{-3}$  & 35.2$\times 10^{-3}$ \\

   \hline \hline 
 \end{tabular}
 \caption{Drag-thrust decomposition of free swimming black ghost knifefish fin, eel, larval zebrafish. Also shown is 
the decomposition of a hypothetical fin with mackerel kinematics. The amplitude shown for the knifefish is at distance of $0.75$~cm 
from base of the fin. The range for $x$ in $A(x)$ is $[0,1]$. Undecomposed forces equal to zero indicate free swimming cases. 
}
 \label{tab:angzebra}  
\end{table}

The drag, thrust, and undecomposed forces on a black ghost knifefish (gymnotiform), an eel (anguilliform), a larval zebrafish (sub--carangiform), and a mackerel (carangiform) are tabulated in Table \ref{tab:angzebra}. Table \ref{tab:angzebra} shows that the error in decomposing forces on free swimming knifefish, eel, and zebrafish is small compared to that in decomposing  the force on a mackerel. During steady free swimming, the average force in the swimming direction is zero. For a steady free swimming organism, the drag and thrust components must be equal. For the black ghost knifefish and the larval zebrafish, it is seen that the drag and thrust forces are equal to each other within 5$\%$. Note that the knifefish displays a small variation in amplitude whereas the zebrafish has a subcarangiform amplitude variation. For the eel, which has an anguilliform amplitude variation, the drag and thrust forces are equal with an error of $17\%$ (Table \ref{tab:angzebra}). Thus, we see that our decomposition works well for gymnotiform (and by similarity for balistiform and rajiform), anguilliform, and subcarangiform swimmers, none of which display rapid variations in amplitude. But, the decomposition does not work well for carangiform swimmers because the rate of change of amplitude in these swimmers is very high (Table \ref{tab:angzebra}). A theoretical assessment of why the decomposition works well for modest variations
in body or fin amplitude along the body length, but not for large variations in 
amplitude, is presented
in the Discussion Section. 


\subsubsection{Application to a robotic knifefish}
\label{sec:robot}
Here we examine how well the 
decomposition works at moderately high Reynolds numbers. We consider
parameters for a robotic knifefish, approximately three times longer than the 
adult live knifefish. At typical kinematic parameters,
such as two waves along the fin
undulating at $2.5$~Hz, the Reynolds number based on fin length is around $130,000$. 
Decomposition of drag and thrust forces on ribbon-fins at the scale of the robot was 
tested using simulations and experiments with the robot. 
The fin has the same dimensions and kinematic parameters in the
simulation and the experiment. However, the surface of the experimental
fin departs from the simulated fin in a manner that appears to affect our
results, as will be described later.

Simulations of the decomposition were carried out for a fin the same size
as that on the robot, but without a body. In section \ref{sec:optH} and Fig.~\ref{fig:momenplot} we show that the forces on the body and the fin are decoupled, and the presence of the body has no influence on the fin forces. Hence, the decomposition can be carried out with or without the body. We choose to carry out the decomposition without the body to reduce the computational cost. 
We consider a case 
where $U=0$~cm/s and $U_w= 40.75$~cm/s. The net force generated by the fin was found to be $384.7$~mN. In the drag causing 
slithering mode, the fin translated forward at $U=U_{w}= 40.75$~cm/s. 
In the thrust-causing frozen mode the 
fin translated backward with a velocity of $U_{w}-U=40.75$~cm/s (slithering and frozen velocity are 
same because $U=0$~cm/s in the undecomposed mode). Thrust and drag forces were found to be  $427.1$~mN and 
$33.54$~mN, respectively. The difference between thrust and drag force ($393.46$~mN), matches well with the force of the undecomposed mode ($384.7$~mN). 
Based on simulations, we infer that the drag-thrust decomposition is valid at length scales of robotic knifefish's ribbon-fin.

As noted before, the parameters used in the experiments were same
as those used in simulations, above. Forces in the undecomposed,
slithering (drag), and frozen (thrust) modes were measured to be
$226.8$~mN, $229.5$~mN, and $283.3$~mN respectively. 
The difference between drag and thrust force is not equal to the
undecomposed force. 
The undecomposed force and thrust force 
are of the same order of magnitude just as it is in simulations.
This raises two important questions:
i) what is the source of the larger than expected slithering (drag) mode force? ii) given this disagreement, is there a resolution? 
These questions will be addressed in the Discussion Section.

\subsection{Utility of the drag--thrust decomposition}

\subsubsection{Optimal height of a knifefish ribbon fin}
What is the utility of separating drag and thrust in the manner we have proposed? Next we
show that this decomposition provides a powerful predictive tool. To that end, we consider
a specific example problem: given the body of a knifefish,
which is held nearly rigid, what should be the height of its ribbon
fin? 
We also show how drag--thrust
decomposition leads to models that can predict the swimming velocity
of an organism.

\begin{figure}
\centerline{\includegraphics[width=.49\textwidth]{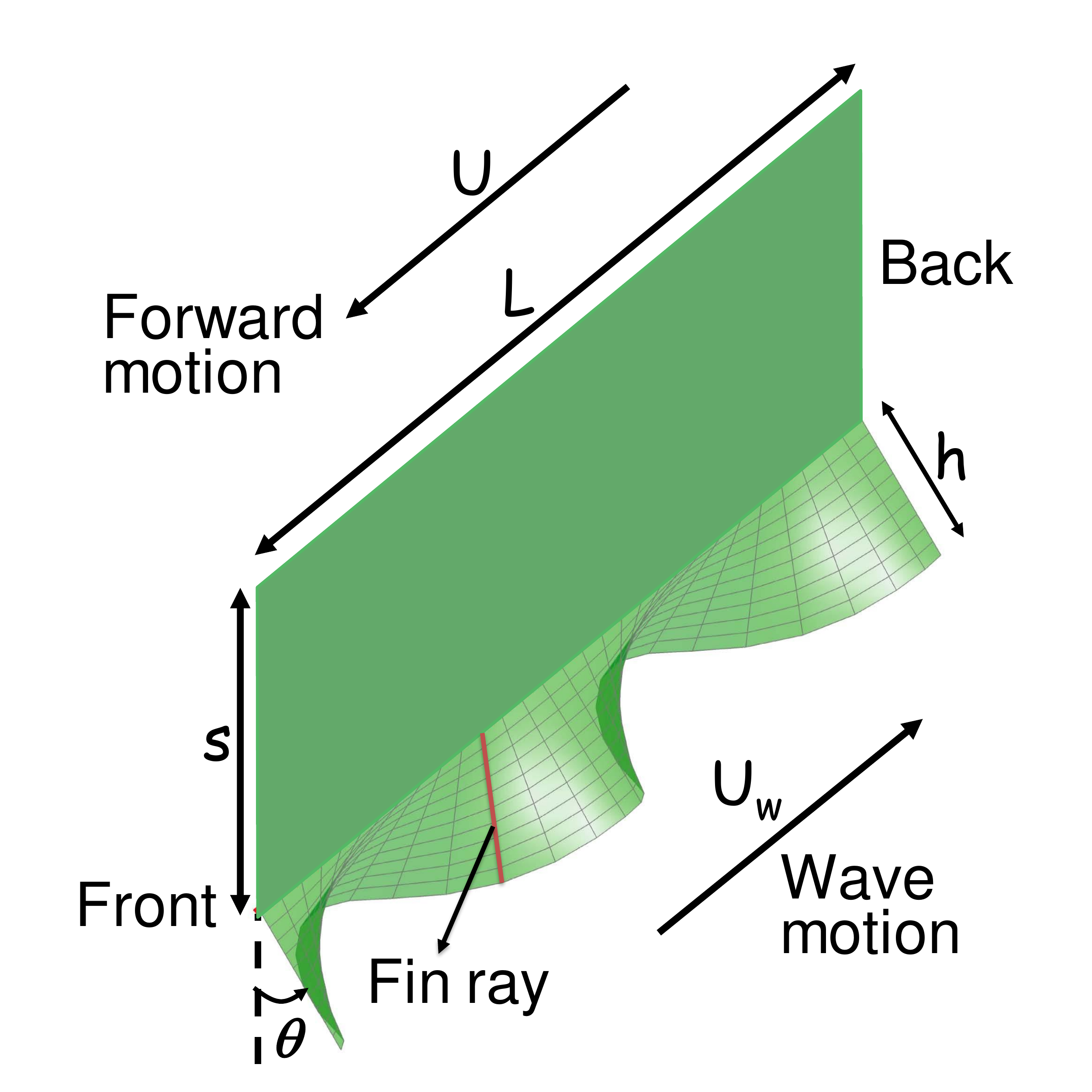}}
\caption{Geometric parameters and the configuration of the
plate-fin assembly.} \label{fig:probsetup}
\end{figure}

To find the preferred fin height, we hypothesized that the observed height of the ribbon fin is such that the mechanical energy spent per unit
distance traveled, referred to as the mechanical cost of transport (COT),
is minimized. The COT was computed numerically for different fin
heights as discussed below. We considered steady swimming in which
a fish moves with a constant mean velocity. For simplicity, we
considered a plate--fin configuration like that used by Lighthill
and Blake \cite{Ligh90a} to study gymnotiform and balistiform
swimming. 
A plate of height $s = 2$~cm
and length $L = 10$~cm was  attached to a ribbon fin of the same
length (Fig.~\ref{fig:probsetup}). These dimensions were selected based on typical
fin and body heights in adult knifefish \cite{Ruiz13a,MacI01a,Blak83a}. 
The following kinematic parameters 
were chosen: $\theta_{max}=30^\circ$, $f=3$~Hz, $\lambda=5$~cm.  
The fin height was varied from $0.5$~cm to $2.5$~cm. For each fin height
we solved the problem of self-propulsion by using a previously
developed efficient algorithm \cite{Shir09a}. 
In these computations the traveling wave motion of the ribbon fin attached to the plate was specified. 
For each case we computed the mean power $P$
spent by the fin against the fluid over one period of the steady swimming
cycle. The time-averaged
swimming velocity $U_{s}$ was estimated during steady
swimming. The cost of transport was computed as COT $= P/U_{s}$.

\begin{figure}
\centerline{\includegraphics[width=.98\textwidth]{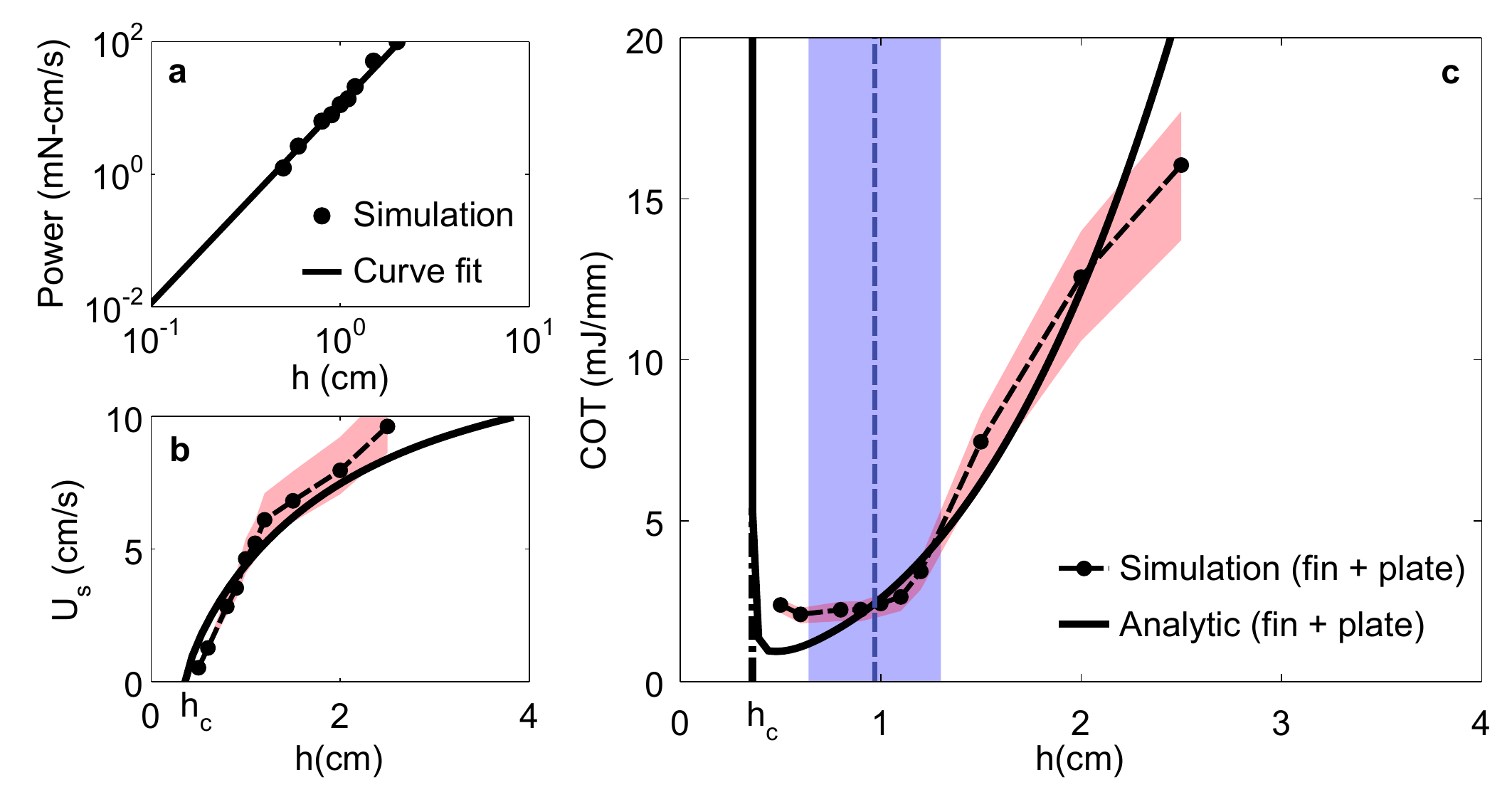}}
\caption{ a) Mechanical power expended in the fluid, obtained from
fully resolved simulations of self--propulsion, as a function of the fin height. b) 
 Swimming velocity as a function of the fin height $h$
obtained from fully resolved simulation of self--propulsion as
well as from a reduced order model (solid line). 
c) The mechanical cost of transport as a function of the fin height $h$ obtained from fully 
resolved simulation of self--propulsion as well as from a reduced order model. The red shaded region represents the variation 
of swimming velocity in (b) and cost of transport in (c) due to a perturbation to the plate height. The perturbation to the height is equal to $\pm 0.85$~cm, which corresponds to the standard deviation 
of the measured body heights in an assortment of knifefish at the half
way point along the fin (see Table~\ref{tab:biodata}). 
The dashed blue vertical line in (c) 
corresponds to the mean fin height measured, and the blue shaded region represents the standard deviation in fin height.
Note: The closed circle and the solid line in (b) and (c) have the same meaning.} 
\label{fig:UsPl}
\end{figure}

Fig.~\ref{fig:UsPl} shows plots of the swimming velocity $U_{s}$, the mean power $P$,
and COT as a function of the ribbon fin height $h$. The power spent on the fluid follows a power law trend (Fig.~\ref{fig:UsPl}a). 
The swimming velocity $U_{s}$ first increases rapidly with respect
to $h$ and then changes slowly at higher values of $h$
(Fig.~\ref{fig:UsPl}b). This trend is a direct result of different
scalings of drag and thrust forces with respect to $h$ 
(Section~\ref{sec:optH}). The COT is low
and nearly constant at smaller $h$ after which it grows rapidly
(Fig.~\ref{fig:UsPl}c). The basis of this increase in COT is that at larger $h$, the power
increases with increasing $h$ but the corresponding increase
in $U_{s}$ is small. Hence there is a rapid growth of COT
at larger $h$. Fig.~\ref{fig:UsPl}c shows that the ribbon fin heights
that give
lower values of COT, for a plate height of $2$~cm, are in the range
of $0.5-1.1$~cm. In this range the COT does not change significantly
but the swimming velocity is highest at $h = 1.1$~cm. 
In short,
different scalings of drag and thrust with respect to $h$ lead to
a specific trend of $U_{s}$ vs. $h$, which in turn determines the
trend of COT vs. $h$. The COT trend eventually provides the prediction
for the fin height $h$ that will minimize the metabolic cost of movement, and as we will
see in the next section, this
predicted height agrees well with observed fin heights. 
\newline
\textbf{\emph{Sensitivity of the optimal fin height to fish body size:}}
The predicted fin height ($\sim 1$~cm) is consistent with the mean
fin height of $0.97$~cm that we measured for $13$ species in $8$
genera in the family Apteronotidae of weakly electric South American knifefishes 
(Table~\ref{tab:biodata}, at 50\% body length). 
The standard deviation of the fin height
from the measured mean value (blue vertical bar in Fig.~~\ref{fig:UsPl}c) is within
the range of fin heights ($0.5-1.1$~cm) for which COT is predicted
to be low. Although the body height of the fishes we considered did
vary, we found from a sensitivity analysis (see Section \ref{sec:pert})
that the influence of the plate (or body) height on the COT trend
is not significant. We show this result in Fig.~\ref{fig:UsPl}c,
where the red shaded region shows the variation in COT due to a change in
plate height corresponding to the standard deviation in the body height
of the 13 species we measured.

\subsubsection{Prediction of swimming velocity}
Finally, to show that the proposed drag-thrust decomposition can be used 
to predict swimming velocities, a force balance equation similar to Eqn. 
\ref{eqn:decomp} was written for the steadily swimming fin--plate assembly 
(Eqn. \ref{eqn:swim-gen}).
We used that equation to derive 
an analytic solution for the fin height as a function of 
swimming velocity (Eqn. \ref{eqn:analy-swim}), which can be rearranged 
to give swimming velocity as a function of fin height. To 
test the analytic prediction of swimming velocity we performed numerical 
simulations to compute the swimming velocity. 
Excellent agreement between the analytic and numerical solutions of swimming velocity 
is shown in Fig.~\ref{fig:UsPl}b.

\section{Discussion}

\subsection{Are there other ways to obtain drag--thrust decomposition?}
\label{sec:OtherDT}
There can be many kinematically consistent decompositions which 
satisfy kinematic conditions D1 
(body movements creating drag summed to body movements creating thrust
result in the original undecomposed movement) 
and D2 (body movements creating drag and thrust are without
discontinuities and physically realizable). However, the kinematic
decomposition depicted in Fig.~\ref{fig:kd} 
is the only
one we have found satisfying D1 and D2 as well as providing force decomposition
(D3, the sum of the decomposed drag and thrust forces is equal to force 
in the originally observed swimming motion of the animal). 
The primary reason that force
decomposition becomes possible is due to boundary layer flow in
the drag mechanism and separated flow in the thrust mechanism (Fig.~\ref{fig:velpress}). 
A boundary layer flow is observed in the drag mechanism because in
that case the velocity on the fin surface is tangential to the
surface itself (Fig.~\ref{fig:velpress}). This type of internal
boundary condition arises because the forward translational
velocity of the fin in the slithering mode is equal and opposite to the wave velocity of
the undulating fin. 
Any translational
velocity in the drag mode that is other than $U_{w}$ will result
in a velocity at each point on the fin that is no longer
tangential to the surface. This gives rise to a flow field
that is not purely due to the boundary layer and it couples with
the separated flow field due to the thrust mechanism. 
Thus, even
if other decompositions are kinematically correct, the force
decomposition will not work well. The kinematic decomposition that
we have is the one that leads to the least
coupling between the drag and thrust modes for the kinematics
considered here.

\begin{figure} 
\centerline{\includegraphics[width=.49\textwidth]{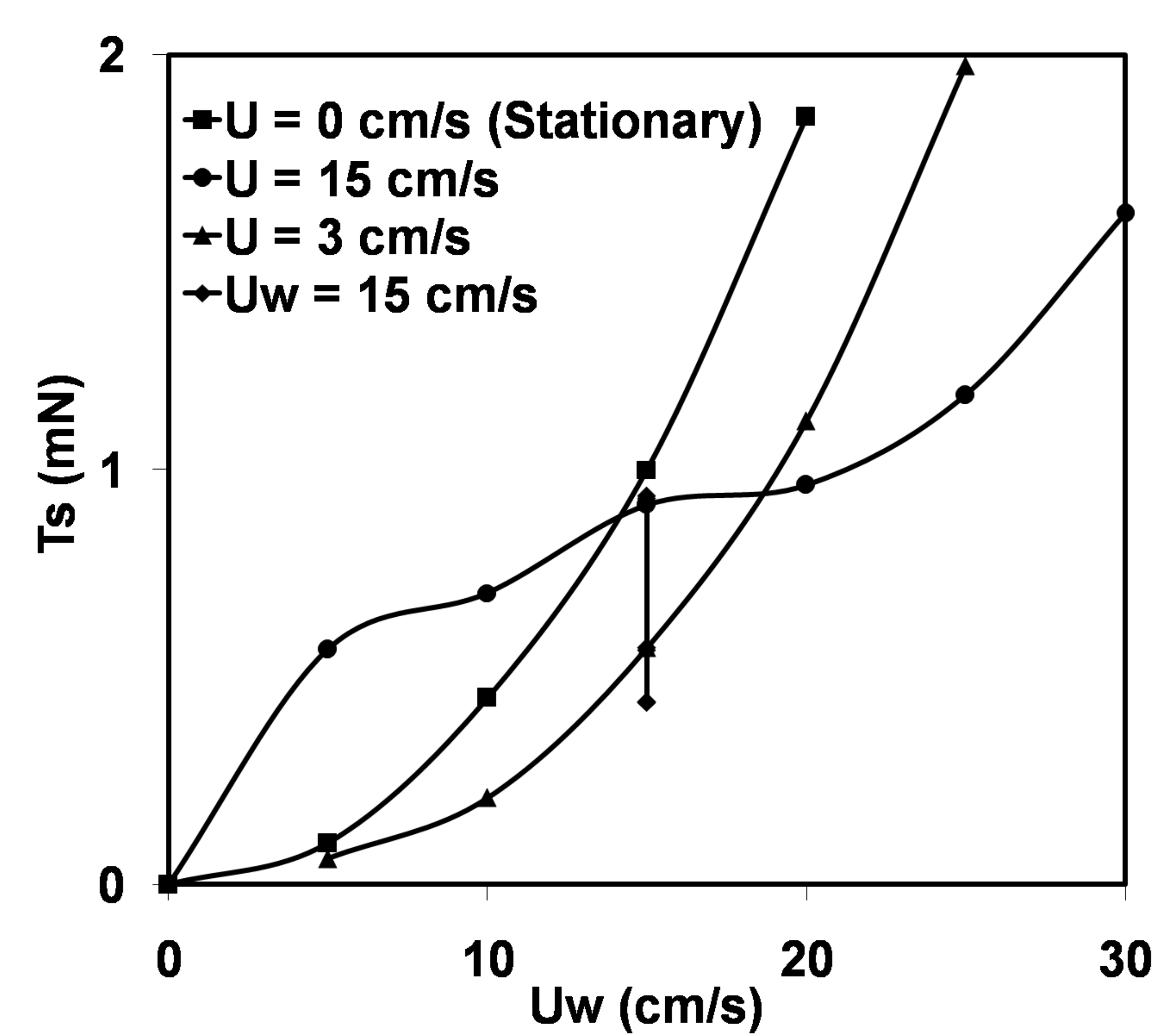}}
\caption{The thrust force $T_{s}$ computed as a function of
$U_{w}$ for each data point in Fig.~\ref{fig:kd} by assuming
a decomposition according to equation \ref{eqn:altsplit}. In this
case the data do not cluster along a single curve. The legend identifies different simulation sets. For example, 
the set with $U_{w}=15$~cm/s was the one where $U_{w}$ was fixed and the value of 
$U$ was changed.}
\label{fig:altsplit}
\end{figure}

As an example, consider a different kinematic decomposition where the
backward traveling wave is \emph{defined} as the thrust producing mechanism whereas
the fin translating forward with velocity $U$ is \emph{defined} as the drag
producing mechanism. We used our data to check if this kinematic
decomposition also leads to force decomposition. If valid then it
should be possible to split the total force as
\begin{equation}
F[U, U_{w}] = T_{s}[U_{w}] - D_{f}[U],
\label{eqn:altsplit}
\end{equation}
where $T_{s}[U_{w}]$ is the forward force on a \emph{stationary}
fin with traveling waves moving backward with wave velocity
$U_{w}$, and $D_{f}[U]$ is the force on a fin with a fixed shape,
i.e. the \emph{frozen fin}, that is translating forward with
velocity $U$. Using the data for $F$ in Fig.~\ref{fig:forces}a 
and the data for $D_{f}[U]$, available from our frozen fin
simulations, we computed $T_{s}[U_{w}]$. These
values are plotted in Fig.~\ref{fig:altsplit} and compared to
the force on a stationary ribbon-fin from our prior work
\cite{Shir08a}. If this decomposition is valid, all data should
fall on a single curve in Fig.~\ref{fig:altsplit}. That is not
the case. It can be shown that the reason this decomposition does
not work is because the corresponding flow fields are not
decoupled.

Given that drag and thrust appear intermingled in swimming
organisms, especially in the undulatory mode, it has been
hypothesized that there must be some spatial or temporal
separation between thrust and drag production that allows the
total force to be zero on average over a swimming cycle
\cite{Tyte04a}. A model to estimate thrust based on temporal
oscillations of the swimming velocity has been proposed
\cite{Peng08a}. Our data suggests that a decomposition of
the total force into drag and thrust is possible without relying
on spatio-temporal splitting.

\subsection{Comments on different measures of drag reported in literature}
\label{sec:DiffDrag}
Appropriate measures of drag on a swimming organism have been
debated in literature for many decades \cite{Fish06a,Schu02a}. Here, we discuss 
how drag from our decomposition is differs from definitions of drag in the literature.
One measure that has been used is the tow-drag, i.e., the drag on a
non-swimming organism if it is pulled in the fluid at its swimming
velocity. The organism is usually not deformed in these
experiments or theoretical estimates. This drag measure is not
expected to be correct because the shape of the animal for the tow-drag estimate 
does not match with shape of the fish during propulsive movement
\cite{Fish06a}. The second measure is the drag obtained by pulling
a deformed non-undulating body through the fluid at its swimming
velocity, i.e., the drag on the frozen shape configuration. As
noted above and in Fig.~\ref{fig:altsplit}, this does not result in
successful decoupling of thrust from drag.
Our results suggest that an appropriate drag measure for undulatory
propulsion is the one corresponding to the perfect slithering
motion at the wave velocity (Fig.~\ref{fig:velpress}).

These measures are best illustrated by considering a hypothetical
swimming ribbon-fin with no body attached to it. According to the
drag-thrust decomposition and the data in Figures \ref{fig:forces}b
and \ref{fig:forces}c, a ribbon-fin with $U_{w} \approx 22$~cm/s
will swim with a velocity $U \approx 9.5$~cm/s. The first drag
measure - the tow-drag - for this case corresponds to the drag on
a flat plate towed at $9.5$~cm/s. This is estimated to be
$0.12$~mN based on boundary layer theory. The second drag measure
corresponding to a frozen fin, moving at $9.5$~cm/s, is $0.6$~mN.
Finally, the third drag measure proposed by us corresponding to
the perfect slithering motion with $U_{w} \approx 22$~cm/s is
$1$~mN. Thus, our drag measure is higher than the other two
estimates for the scenario considered here. The result is
consistent with reports in literature that the tow-drag is often
found to be lower than that required to achieve a balance of drag
and thrust forces during swimming \cite{Fish06a,Schu02a}. In
general, however, the relative magnitudes of the three drag
measures may not be in the same order as in the example discussed
above. It will depend on various parameters including the
geometric configuration.

It has been noted in the past that body undulations lead to a
reduction in drag on a swimming body \cite{Shen03a}. That
conclusion was based on computing the total force on an infinite
two-dimensional wavy surface for a given imposed velocity $U$ and
then noting that as $U_{w}$ is increased the total force changes
from being backward (drag-like) to being forward (thrust-like). In
this sense the presence of undulations reduces the drag-like
behavior. This is consistent with our results.

\subsection{Limits of the drag--thrust decomposition}
\label{sec:GenDecomp}
Results of the decomposition of forces on swimming animals showed
that the decomposition is valid for swimming animals with anguilliform,
gymnotiform (by similarity balistiform and rajiform), and subcarangiform
kinematics. But, the decomposition is not valid for swimming animals
with carangiform kinematics. Here, we present a theoretical assessment
of when the proposed decomposition will be correct and when it will fail. 
For the purpose
of analysis, consider kinematics imposed on a rectangular surface
(``fin'' hereafter) of infinitesimal thickness. At any
instant a given point on the fin undergoes lateral displacement
given by
\begin{equation}
y=A(x)\sin[2\pi(\frac{x}{\lambda}+ft)],
\label{eqn:tw}
\end{equation}
where $A(x)$ is the amplitude which is a function of axial direction and it determines the mode of swimming. 
The amplitude for different modes of swimming is shown in Table \ref{tab:angzebra}. The 
deviation from exact decomposition is a function of the rate of increase of amplitude with  body length. The 
deviation will be more if the rate of amplitude change is high and vice-versa. This  will be demonstrated below. 

The slithering mode is affected by a deviation from exact decomposition. The slithering mode has the 
property that velocity at each point on the fin is tangential to surface of the fin. The velocity in the slithering 
 mode, at a point, is the resultant of the forward translational velocity $U_{w}$ and the lateral velocity $V_{w}$. 
  The angle, $\alpha$ (Fig.~\ref{fig:kd}), of the velocity at a point is given by
\begin{equation}
\tan\alpha=\frac{2\pi A(x)\cos[2\pi(\frac{x}{\lambda}+ft)]}{\lambda}.
\label{eqn:tantheta}
\end{equation}
The direction of the resultant velocity at every point on the fin surface must be equal to slope of 
the corresponding point if the velocity has to be tangential to the fin surface. The slope of a point on the fin
undergoing  traveling wave motion is given by
\begin{equation}
\frac{dy}{dx}=\frac{2\pi A(x)\cos[2\pi(\frac{x}{\lambda}+ft)]}{\lambda}+\frac{dA(x)}{dx}\sin[2\pi(\frac{x}{\lambda}+ft)].
\label{eqn:slope}
\end{equation}
Note that if $A(x)$ is constant then  equation \ref{eqn:tantheta} and \ref{eqn:slope}  are 
identical resulting in a perfect slithering motion where the resultant velocity of a point is tangential 
to the fin surface. Substituting equation \ref{eqn:tantheta} in 
\ref{eqn:slope} we get the following equation
\begin{equation}
\frac{dy}{dx}-\tan\alpha=\frac{dA(x)}{dx}\sin[2\pi(\frac{x}{\lambda}+ft)].
\label{eqn:tan-slope}
\end{equation}
The above equation is a measure of deviation from exact decomposition. At a given instant of time, the 
deviation from tangential velocity is directly proportional to $\frac{dA(x)}{dx}$. The figures in the first 
row of Table \ref{tab:angzebra} show that the rate of amplitude change in anguilliform swimmers, subcarangiform swimmers and knifefish is slower than that in 
carangiform swimmers. In anguilliform swimmers the amplitude gradually increases from head to tail, i.e, the rate of change of 
amplitude is smaller. Thus, the deviation from perfect slithering
motion is small. The same is true for the subcarangiform swimmer
as well. In case of the knifefish amplitude, the amplitude first
increases, reaches a peak, and then decreases. But the rate at
which amplitude increases or decreases is very small, hence the
deviation from perfect slithering motion is small. Thus, the
decomposition of eel, larval zebrafish and knifefish worked well.
However, in carangiform swimmers, the
amplitude remains relatively small and constant for at least half the body length after which it increases 
rapidly. The rate of change of amplitude is very high towards the caudal portion of the body thus resulting in a 
large deviation from perfect slithering motion. Consequently, it is not 
surprising that the decomposition did not work for mackerel.

\begin{figure} 
\centerline{\includegraphics[width=.49\textwidth]{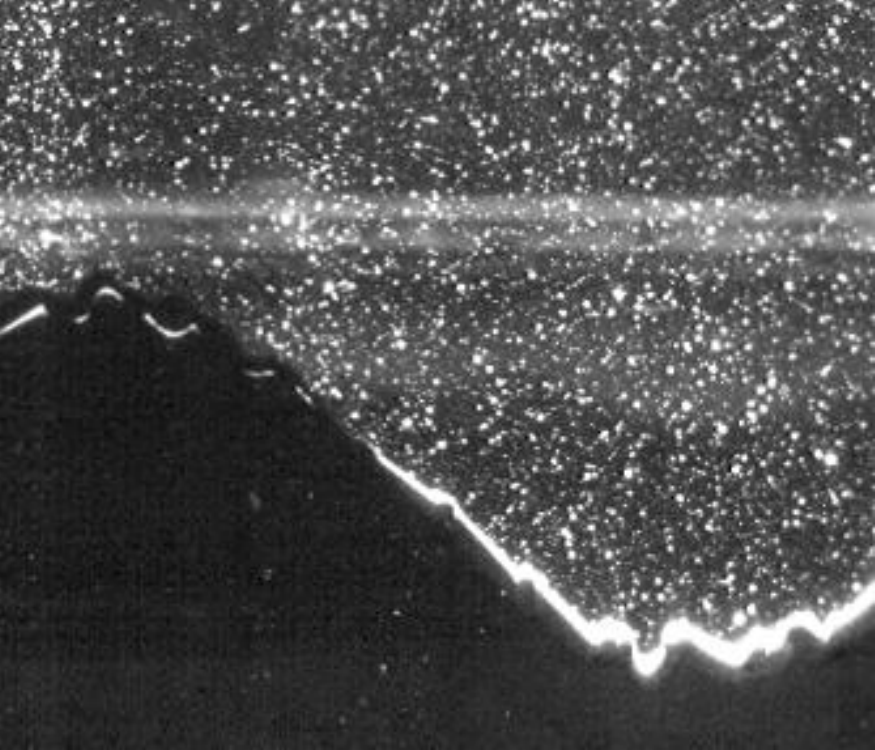}}
\caption{The surface of the robotic ribbon fin during sinusoidal undulations. The material between adjacent rays folds and results in kinks on the fin surface.} 
\label{fig:robotfin}
\end{figure}

\subsection{Why the drag--thrust decomposition failed for the robotic knifefish}
\label{sec:robot-discuss}
\begin{figure} 
\centerline{\includegraphics[width=.49\textwidth]{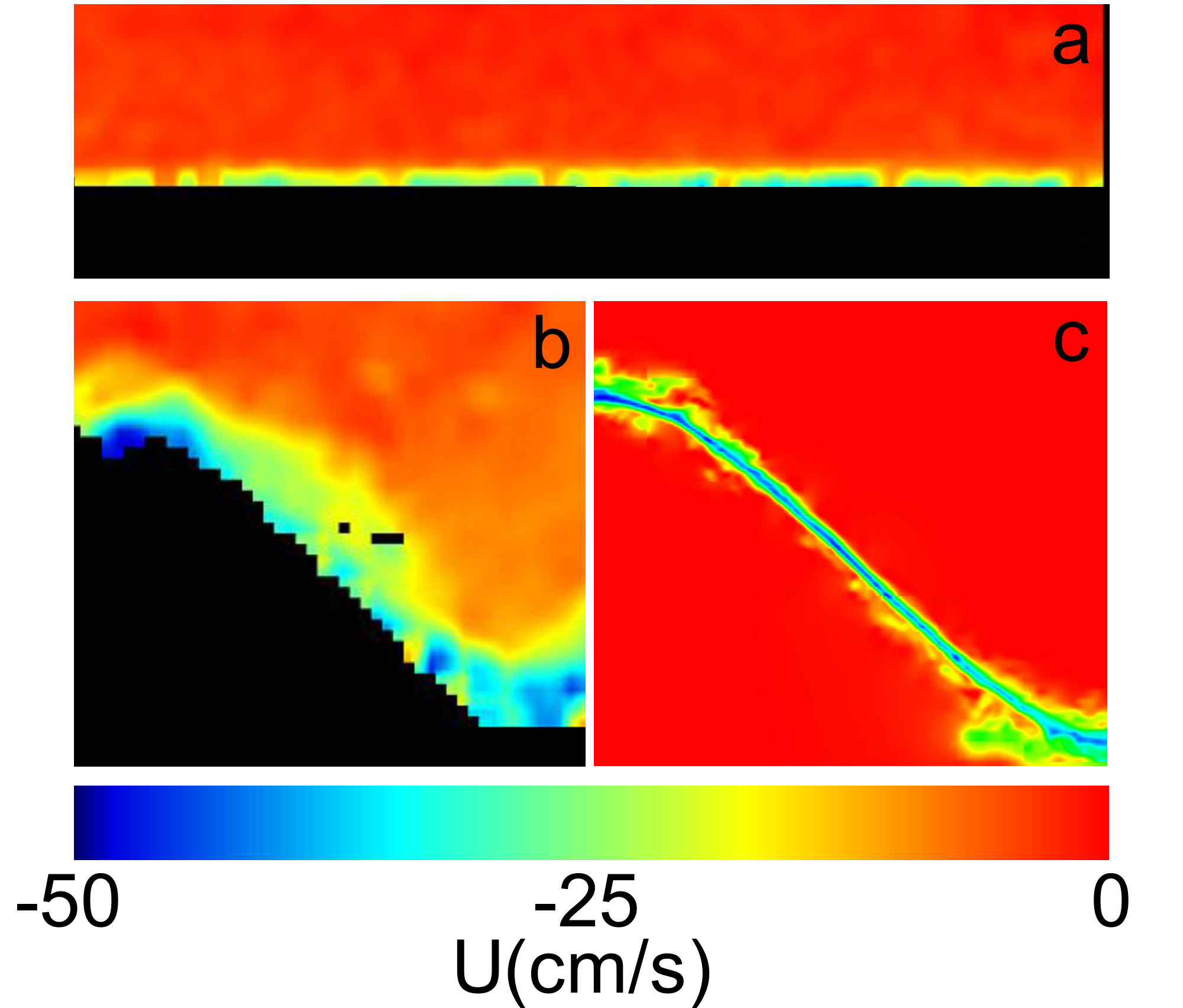}}
\caption{A comparison between the boundary layer due to,  a) flow past a flat plate, 
b) slithering motion of a robotic ribbon fin, c) and slithering motion of a modeled robotic fin in a simulation. The boundary layer is 
shown on a horizontal plane 1~cm above the bottom edge of the fin 
toward the robot body. The color bar represents the axial velocity.} 
\label{fig:BoundLayer}
\end{figure}

Using numerical simulations it was shown that the decomposition was valid for the robotic knifefish. But, experimental data did not agree with the simulations. The main difference between the experiments and the simulation was in the force of the slithering mode. The slithering drag force from experiment was higher than that from simulation; this implies that the drag force is higher than what it should be for the decomposition to be valid. We hypothesize that the larger than expected slithering drag force in the experiment is due to the imperfections on the robot's fin surface. The robotic fin is made up of discrete rays ($32$ rays $1$~cm apart) that are connected 
by Lycra fabric; see Fig.~\ref{fig:Robot} in which one of the fin rays is highlighted in white. It is not 
possible to produce a smooth sinusoidal wave on the fin unless the fin is made up of very large number of rays. 
Owing to the limited number of rays, the sinusoidal wave generated by the robotic fin is not smooth (see Fig.~\ref{fig:robotfin}). 
With kinks on its surface the robotic fin cannot maintain a thin boundary layer in the slithering mode like that in simulations. The 
kinks will introduce disturbances into the boundary layer (see Fig.~\ref{fig:BoundLayer}). 
Unlike its real counterpart, the robotic fin modeled in the simulation is composed of very fine grid points 
(whose resolution is of the order of fluid grid resolution). Any curved surface can be imposed (sinusoidal or 
otherwise) on the modeled fin surface without causing any kinks. Hence, it leads to a thin undisturbed boundary 
layer in the slithering mode (see Figure  \ref{fig:BoundLayer}). The boundary layer on the robotic fin is very thick when compared to that from simulation or that from flow past a flat plate  at the same Reynolds number (see Figure  \ref{fig:BoundLayer}). 
The large thickness of the boundary layer may also indicate separated flow, which could lead to the large drag force measured in the experiment. 

\subsubsection{An alternate estimate of the drag force on the robotic ribbon fin}
Given that the slithering mode force in the experiment does not satisfy the drag--thrust decomposition due 
to disturbances caused by surface kinks, we propose that the next best choice may be to use a drag estimate 
that is similar to the tow--drag. This is because the surface imperfections on an undeformed (or straight)
robotic ribbon--fin would be much less and consequently the flow would be less perturbed or separated. There is subtle, yet important, difference between our estimate of drag and the conventional measure of tow-drag. In the conventional measure of tow--drag, the swimmer is towed at its swimming speed in stationary water. In contrast, we measure the drag by towing a straight fin, in stationary water, at the velocity of the slithering mode (which is equal to wave velocity, $U_{w} = 40.75$~cm/s) instead of swimming speed. The force was measured to be $81.9$~mN. This value is closer to 
the expected value of $56.6$~mN obtained from the drag thrust decomposition estimate.

\subsubsection{Why the drag--thrust decomposition works for real knifefishes}
Simulations have shown that the decomposition is indeed valid at moderately high Reynolds number. Experiments 
on the robotic knifefish, however,  suggest possible limitations of the decomposition. 
The decomposition is sensitive to any deviation from perfect slithering mode, 
be it due to high levels of amplitude
change or high roughness on the fin surface. 
A knifefish's ribbon-fin
is composed of 120--320 rays spanning fin 
lengths of 10--30~cm, giving typical densities of about 
one ray per millimeter (Table 4 and Fig.~41 of \cite{Albe01a}). Many of these
fin rays branch into two rami half the way from their base to their distal ends, 
which would effectively double the ray density along the very portion of the fin
surface that could become uneven due to spreading between the rays when they
are oscillated \cite{Albe01a}.  
With such fine ray spacing, a knifefish can produce curved surfaces on its 
fin without kinks.  In contrast,
on the robot the ray density is one ray per 10~mm, resulting in a much less
smooth fin surface.
We therefore expect that drag-thrust decomposition would be applicable 
to a robotic ribbon-fin if it can be designed to have higher ray density
so that the fin surface is smoother. 

\section{Methods}
\label{sec:MatMeth}
\subsection{Experimental setup}
\subsubsection{Robotic undulating fin}
\begin{figure}
\centerline{\includegraphics[width=.49\textwidth]{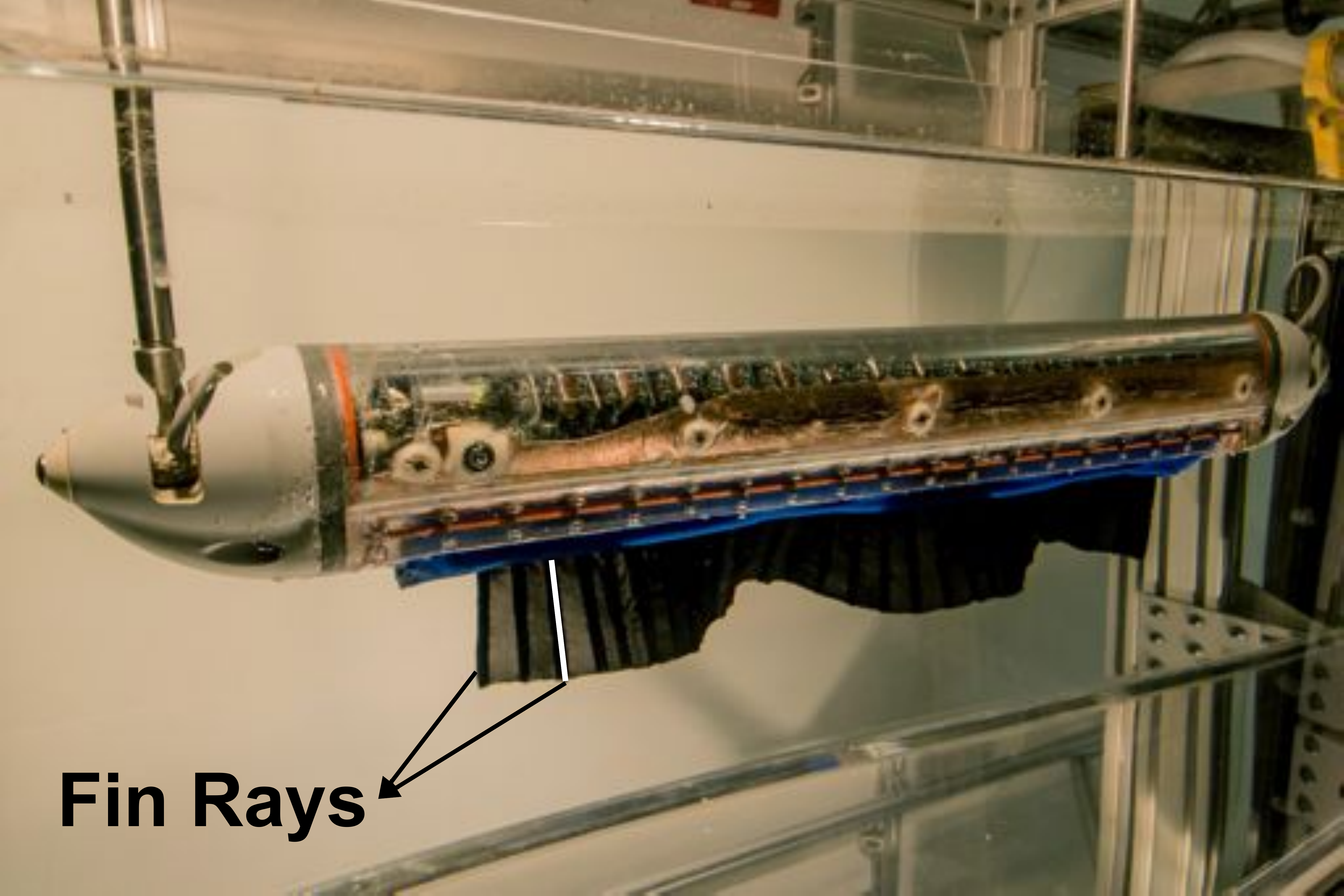}}
\caption{The robotic knifefish used in the drag-thrust decomposition experiments. The robotic ribbon-fin is composed of 32 fin rays and a Lycra 
fabric  connecting the rays. One of the fin rays is coloured white to highlight it.} 
\label{fig:Robot}
\end{figure}

The robotic model used for the experimental work was the \textquoteleft Ghostbot,\textquoteright\ a biomimetic knifefish robot which undulates an elongated fin to generate thrust (see Fig.~\ref{fig:Robot}).  The robot consists of a rigid cylindrical body that houses the motors and electronics to drive the individual rays of the fin.  There are 32 rays to actuate a rectangular Lycra fin measuring 32.6~cm by 5~cm.  More details of the robot are found in \cite{Cure11a}, with the only difference being the depth of the fin was 3.37~cm in the previous work rather than 5~cm.

\subsubsection{Measuring hydrodynamic forces}
The robot was suspended horizontally into a variable speed flow tank from an air-bearing platform allowing near frictionless motion in the longitudinal axis.  We fixed the robot in the lateral and vertical directions.  We placed a single axis force transducer (LSB200, Futek, Irvine, CA, USA) in the longitudinal axis between the air-bearing platform and mechanical ground, allowing us to measure the forces generated by the robot or acting on the robot along that axis.  Voltages from the force transducer were recorded at 1000~Hz.  For each trial, we allowed ample time for the hydrodynamics to reach steady state (30-60 seconds), then averaged the last 10 seconds of data, which was converted to force units based on the calibration, which had a maximum nonlinear error of 0.034\%.  The flow speed of the water tunnel was measured and calibrated using particle image velocimetry (PIV).  More details on PIV are provided in the following section.

\subsubsection{Particle image velocimetry (PIV)}
We analyzed horizontal PIV planes to measure the boundary layer thickness caused by the fin.  The PIV setup used is the same as the one described in \cite{Neve13b}.  In short, a 2~W laser beam (Verdi G2, Coherent Inc., Santa Clara, CA, USA) is scanned at 500~Hz to create a planar laser light sheet.  A high-speed camera (FastCam 1024P PCI, Photron, San Diego, CA, USA) imaged reflective particles suspended in the fluid (44 micron silver coated glass spheres, Potter Industries, Valley Forge, PA, USA) at 500 frames per second, matching the scanning rate of the laser.  High-speed video was analyzed using a commercial software package (DaVis, LaVision GMBH., G\"{o}ttingen, Germany).  Successive frames of the video were cross-correlated to calculate the velocity vector field of the fluid.  Cross-correlation consisted of two passes with decreasing interrogation windows, first with an interrogation window of 32 by 32 pixels with 50\% overlap and second with a window of 16 by 16 pixels with 50\% overlap.

\subsection{Numerical problem formulation}
For the numerical simulations, the ribbon fin is modeled as a thin membrane as shown in
Fig.~\ref{fig:gymbali}. The angular position $\theta(x,t)$ of
any point on the fin (described above in under Results) is modeled as
\begin{equation}\label{eq:travwav}
    \theta(x,t) = \theta_{max} \sin2\pi(\frac{x}{\lambda}- f t).
\end{equation}
This corresponds to a sinusoidal traveling
wave along the fin of length $L$ and height $h$. Kinematic
parameters are frequency $f$, $\theta_{max}$, and $\lambda$. The speed 
at which the wave form travels along the fin is called the wave velocity, $U_w = f\lambda$. 
In the optimal ribbon fin height analysis (presented in Results
section), the knifefish is modeled as a plate-fin assembly where a
rigid plate is attached to the ribbon-fin in place of the fish's
rigid body. In these simulations, the rigid plate is modeled as a
rigid surface.  The properties of water
are used for the fluid. Two types of simulations are performed in
this work. In one type, the translational velocity $U$ of the fin
and/or plate is specified along with the deformation kinematics of
the fin (equation \ref{eq:travwav}). These simulations are carried
out in the frame of reference of the fin/plate. Hence the
translational velocity $U$ appears as an imposed free stream
velocity. Rotation of the fin/plate is prohibited. 

In the second
type of simulation, only the deformation kinematics of the fin are
prescribed, which result in a self-propelling fin-plate assembly.
Complete details of the computational method and validation are
given in \cite{Shir08a,Shir09a}. In this method, the viscous
Navier-Stokes equations along with the incompressibility
constraint are solved in the entire domain. The effect of the
immersed fin/plate is resolved by a new constraint based
formulation described in \cite{Shir09a}. A finite difference
method that is $6^{th}$ order in space and $4^{th}$ order in time
is used. The grid size was chosen after performing a
grid-sensitivity study. The Courant-Friedrichs-Lewy (CFL) number
is $0.25$ for all simulations. Periodic
boundary conditions are used in all directions. The computational
domain was made large enough to minimize the impact
of periodicity. Mean forces and power of the fin were calculated
as the time average over at least one period of oscillation, after
a quasi-steady state is reached.

Parameters chosen to match those of an adult black ghost
knifefish \cite{Blak83a} are: fin length $L=10$~cm, fin height $h=1$~cm,
$f=3$~Hz, $\theta_{max}=30^\circ$, and
$\lambda=5$~cm. The density and viscosity of water are
taken as $\rho=1,000$~kg/m$^3$ and $\mu=8.9\times
10^{-4}$~kg/m$\cdot$s unless otherwise specified.

\newcolumntype{M}[1]{>{\centering\arraybackslash}m{#1}}
\renewcommand{\arraystretch}{2}

\begin{table}
\centering
 \begin{tabular}{M{4.5cm}  M{2.5cm}   M{2.5cm}  M{2.5cm}  M{2.5cm} }  \hline \hline
  
                  & \textbf{Knifefish} & \textbf{Eel}  &  \textbf{Zebrafish (larval)} &\textbf{Mackerel} \\ \hline \hline
                                 
   \textbf{Body/Fin dimensions 
  length$\times$height (cm$^2$)}    & Actual (9$\times$1) & 1$\times$0.1 & Actual (0.38$\times$0.04) & 1$\times$0.2 \\

   \textbf{Wavelength $\lambda$ (cm)}  & 3.75    & 1  &  0.31 & 1 \\ 

   \textbf{Frequency $f$ (Hz)}     & 10.3  & 5 & 33.3 & 15.7 \\ 
   
   \textbf{Fluid density (kg/m$^3$)} & 1000 & 1000 & 1000 & 1000 \\
   
   \textbf{Fluid viscosity (Pa s)}  & 0.9$\times$10$^{-3}$ & 1.4$\times$10$^{-5}$& 0.9$\times$10$^{-3}$& 0.9$\times$10$^{-3}$ \\

   \textbf{Wave velocity (cm/s)}  & 38.62 & 5 & 10.32 & 15.7 \\
   
   \textbf{Translational velocity (cm/s)}  & 11.9 & 2.89 & 1.125 & 10 \\
   
   \hline \hline 
 \end{tabular}
 \caption{Simulation parameters.}
 \label{tab:para}  
\end{table}

\subsubsection{Numerical simulations of swimming animals}
Real three dimensional geometries of the bodies or fins were simulated in 
all cases except the mackerel where a sheet-like fin was simulated with 
mackerel-like kinematics. The simulation parameters and the kinematics are 
given in Table~\ref{tab:para}. The fin profile and the kinematic 
data of the knifefish were experimentally obtained by us \cite{Ruiz13a}. 
Only the ribbon-fin of the knifefish was considered for the decomposition. 
The body of the knifefish was not considered for the same reason the 
body was not considered in the robot simulation. 
Experimentally extracted kinematic data 
and body profile of the larval zebrafish were provided by Melina Hale of The 
University of Chicago. The body profile of the eel was taken from Kern and Koumoutsakos' \cite{Kern06a} analysis of 
anguilliform swimming. The kinematics of both the mackerel and eel were described by equation \ref{eqn:tw}, where the amplitude of kinematic 
undulations are based on experiments. Eel kinematics were based on experiments by 
Tytel and Lauder \cite{Tyte04a}, and mackerel kinematics were based on experiments by Videler and Hess\cite{Vide84a}.

\subsubsection{Sign convention}
As shown in Fig.~\ref{fig:probsetup} the forward direction is to
the left and the backward direction is to the right. Translational
velocities $U$ and $U_s$ are positive if directed to the left
(forward) while the wave velocity $U_w$ is positive if directed to
the right (backward). All forces considered in this work are
parallel to the length of the plate and the fin. Thrust force
\emph{on} the fin is positive to the left while drag forces
\emph{on} the fin and plate are positive to the right. The
resultant force \emph{on} the fin due to thrust and drag is
positive to the left.


\newcolumntype{M}[1]{>{\arraybackslash}m{#1}}
\renewcommand{\arraystretch}{1}

\begin{sidewaystable}
\centering

 \begin{tabular}{M{3.2cm}  M{2.3cm}   M{1.1cm}  M{0.9cm}  M{0.9cm}  M{0.9cm} | M{0.8cm}   M{0.8cm}  M{0.8cm}|  M{0.8cm}  M{0.8cm} M{0.8cm}|  M{0.8cm}  M{0.8cm}  M{0.8cm}}  \hline \hline
    &&&&&&&&&&&&&& \\
                 & & & & & & \multicolumn{3}{c|}{25\% FL} & \multicolumn{3}{c|}{ 50\% FL} &  \multicolumn{3}{c}{ 75\% FL} \\ 
                    
 Genus  & Species & Sp. ID & BL & BLF & FL & S & H & R & S & H & R &S & H & R \\     \hline 
         
Adontosternarchus & devanazii & 59522 & 8.30 & 10.00 & 7.20 & 1.10 & 0.60 & 1.83 & 0.86 & 0.65 & 1.32 & 0.45 & 0.50 & 0.90 \\
Adontosternarchus & clarkae & 78151 & 8.30 & 8.30 & 7.40 & 1.20 & 0.60 & 2.00 & 0.92 & 0.63 & 1.47 & 0.98 & 0.43 & 2.27 \\
Adontosternarchus & sachsi & 92914 & 11.70 & 15.80 & 10.50 & 1.40 & 0.62 & 2.26 & 1.12 & 0.68 & 1.65 & 0.66 & 0.50 & 1.32 \\
Apteronotus & leptorhynchus & 48686 & 10.90 & 12.80 & 9.40 & 1.82 & 0.60 & 3.03 & 1.49 & 0.71 & 2.10 & 0.80 & 0.71 & 1.13 \\
Magosternarchus & duccis & 46884 & 11.20 & 13.00 & 10.50 & 1.80 & 0.78 & 2.31 & 1.50 & 0.73 & 2.05 & 0.90 & 0.65 & 1.38 \\
Apteronotus & bonapartii & 78152 & 14.10 & 16.70 & 12.30 & 2.00 & 0.60 & 3.33 & 1.60 & 0.65 & 2.46 & 0.90 & 0.53 & 1.71 \\
Sternarchella & schotti & 59517 & 14.80 & 17.20 & 12.90 & 2.04 & 0.87 & 2.34 & 1.60 & 0.95 & 1.68 & 0.98 & 0.73 & 1.34 \\
Sternarchella & terminalis & 98361 & 14.80 & 15.60 & 12.40 & 2.30 & 1.00 & 2.30 & 1.80 & 1.06 & 1.70 & 1.10 & 0.97 & 1.13 \\
Porotergus & gimbeli & 164198 & 15.70 & 15.70 & 14.50 & 2.10 & 1.06* & 1.98 & 1.85 & 1.05 & 1.76 & 1.25 & 0.83 & 1.51 \\
Apteronotus & albifrons & 78148 & 12.90 & 12.90 & 10.20 & 2.40 & 0.83 & 2.91 & 1.95 & 0.99 & 1.97 & 1.10 & 0.95 & 1.16 \\
Sternarchogiton & porcinum b & 164197 & 17.50 & 20.50 & 14.70 & 2.40 & 1.15 & 2.09 & 2.00 & 1.20 & 1.67 & 1.10 & 0.60* & 1.83 \\
Apteronotus & albifrons & 169184 & 16.40 & 19.50 & 13.50 & 3.10 & 0.93 & 3.35 & 2.40 & 1.10 & 2.18 & 1.40 & 1.10 & 1.27 \\
Sternarchorhamphus & muelleri & 59514 & 33.00 & 39.70 & 26.90 & 3.88 & 1.75 & 2.22 & 2.82 & 1.53 & 1.84 & 1.80 & 1.26 & 1.43 \\
Orthosternarchus & tamandua & 98368 & 35.60 & 35.60 & 28.10 & 4.34 & 1.67* & 2.60 & 4.18 & 1.64 & 2.55 & 2.90 & 1.10 & 2.64 \\

\hline
&&&&&&&&&&&&&& \\
 & & &\multicolumn{3}{c|}{Mean} & 2.28 & 0.93 & 2.47 & 1.86 & 0.97 & 1.89 & 1.17 & 0.78 & 1.50  \\
  &&&&&&&&&&&&&& \\
  & & & \multicolumn{3}{c|}{Standard Deviation} & 0.94 & 0.38 & 0.50 & 0.85 & 0.33 & 0.35 & 0.59 & 0.26 & 0.47 \\ 
 
 \hline\hline
 \end{tabular} 
\captionof{table}{Body length (BL), body length with filament (BLF), fin length (FL), body height (S), fin height (H), ratio
of body height to fin height (R) of 13 species across 8 genera of South American weakly electric fishes in the Apteronotidae, 
in order of body height at 50\% along fin length. Specimen 
identification numbers (Harvard Museum of Comparative Zoology) are represented 
by Sp. ID. Body height, fin height, and their ratio (R) were measured at 25, 50, and 75 percent along 
the fin length, and the same are tabulated. Note: All lengths are in centimeters, asterisk indicates that the quantity in question was 
estimated, not measured.} \label{tab:biodata}

\end{sidewaystable}

\subsection{Measurements of fin and body size of South American weakly electric fishes}
\label{sec:biodata}
A group of 13 species in 8 genera in the family Apteronotidae 
of South American weakly electric fishes 
were considered.  The family Apteronotidae 
is one of five families encompassing 32 genera and
135 species of South American knifefishes \cite{Albe05b}. 
We restricted our measurements to a
subset of one family for practical reasons, but it is evident from illustrations
and images of knifefish in other families that the Apteronotidae are quite typical
in terms of the body height to fin height issue examined here 
\cite{Albe01a,Cram06a,Albe05b}.

\begin{figure} 
\centerline{\includegraphics[width=.99\textwidth]{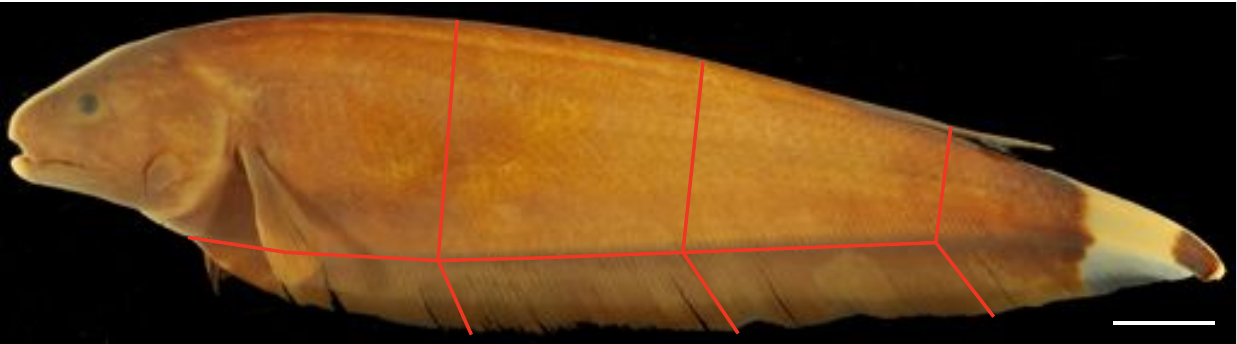}}
\caption{The lines along which the fin length, fin height, and body
height were measured for \emph{Apteronotus albifrons.} Scale bar:
10~mm.}
\label{fig:bio-measure}
\end{figure}

An image of each specimen 
was provided by Andrew Williston of the Museum of Comparative Zoology, 
Harvard University.  
The fin and body height 
measurements were made at three locations along the fin length from the rostral tip: $25$, $50$, and $75$ percent. The height of the fin was measured by 
measuring the length of the collapsed fin ray. The height of the body was measured along a line that was perpendicular 
to the body axis. The lines of measurement are shown in Fig.~\ref{fig:bio-measure}. The measured data are tabulated in 
Table \ref{tab:biodata}. For specimens in which fin rays were
not present for measurements at a needed position along the fin, 
ray length was estimated based on the trend of neighboring fin ray lengths. These 
data points are marked by an asterisk in the table.

\section*{Acknowledgments}
This work was supported by NSF grants CBET-0828749 to N.A.P. and
M.A.M, CMMI-0941674 to M.A.M. and N.A.P, and CBET-1066575 to N.A.P. Computational resources were provided by NSF's TeraGrid Project grants CTS-070056T and CTS-090006, and by Northwestern University High Performance Computing System -- Quest.

\section*{Author contributions} NAP, AAS, RB, MAM, and IDN conceived and designed the research and experiments. RB and AAS performed numerical simulations. IDN and RB performed experiments. NAP, RB, AAS, and MAM analyzed data. AAS, APSB, and NAP contributed the numerical simulation tool. IDN and MAM contributed the experimental tool. NAP, RB, AAS, and MAM wrote the paper. All authors edited the paper.

\newpage

\section*{Appendices}

\appendix

\section{Difference between this work and Lighthill's approach}
\label{sec:OurvsLight}

Our approach is different from prior work by Lighthill \cite{Ligh75a,Ligh71a,Ligh90a} for the
following reasons: \textbf{1)} Our kinematic decomposition ensures
that the realizability condition D2 
(the decomposed body movements should be such that the  surface of the body will move in a continuous fashion so that the kinematics can be realized in experiments or simulations) 
is satisfied, unlike Lighthill's decomposition.
Thus, our decomposition allows setting up independent sets of
simulations/experiments to determine the drag and thrust
components. \textbf{2)} We show that the force condition D3 
(the sum of the drag and thrust forces from the decomposed body movements must be equal to the force due the original body movement of the swimmer) 
is satisfied
based on our approach. Because Lighthill's approach did not
satisfy condition D2, this verification has not been possible.
\textbf{3)} We are able to estimate both thrust and drag, unlike
Lighthill's work where appropriate drag estimates are not
available. Lighthill's approach for the decomposition of drag and 
thrust is summarized below.

\begin{figure}[t!]
\centerline{\includegraphics[width=.49\textwidth]{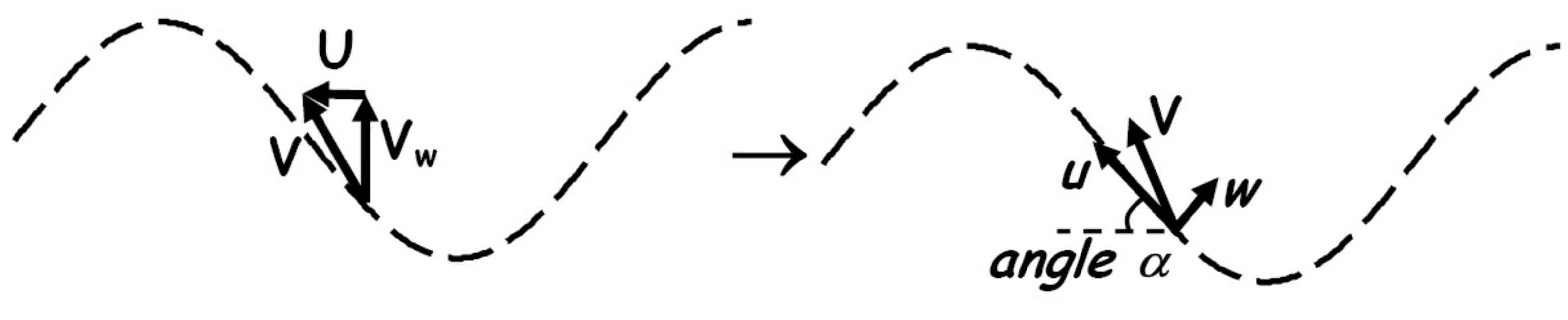}}
\caption{Lighthill \cite{Ligh75a} decomposed the resultant
velocity $V$ at a point on the surface into a component $u$
tangential to it (causing resistive drag) and a component $w$
normal to it (causing reactive thrust).} \label{fig:Ligh}
\end{figure}

Consider a generic surface with a traveling wave as shown in
Fig.~\ref{fig:Ligh}. It could represent the body of a swimmer or
the surface of a fin.  Consider the wave to have a constant
amplitude and that the wave is traveling from left to right (i.e.
backward) with velocity $U_{w}$ which is constant. Due to the wave
motion, any point on the surface oscillates laterally (or
oscillates laterally along a figure eight-shaped path if the body
is assumed inextensible \cite{Ligh75a}) with velocity $V_{w} = -2
\pi f A \sin[2 \pi (x/\lambda - f t)]$, where $A$ is the amplitude
of the wave, $f$ is the frequency, $x$ is the axial location of
the point, and $\lambda$ is the wavelength. It follows that $U_{w}
= \lambda f$. Let $U$ be the forward translational velocity (i.e.
to the left; Fig.~\ref{fig:Ligh}) of the surface as a whole
and let the surrounding fluid be stationary. Positive
values of $U_{w}$ will imply a backward traveling wave whereas
positive values of $U$ imply a forward translating surface. The
resultant velocity at any point on the surface is $V$ as shown in
Fig.~\ref{fig:Ligh}. Lighthill \cite{Ligh71a} considered the
components of $V$ that are tangential ($u$) and normal ($w$) to
the surface of the wave with large amplitude oscillations. It can be
shown that
\begin{equation}
w = (U_{w}-U) \sin[\alpha], \label{eqn:w}
\end{equation}
\begin{equation}
u = \frac{U_{w}}{\cos[\alpha]}-(U_{w} - U) \cos[\alpha],
\label{eqn:u}
\end{equation}
where $\alpha$ is the angle as shown in Fig.~\ref{fig:Ligh}.
Lighthill \cite{Ligh71a} then stated that the added mass due to
the normal component $w$ will be large and therefore it will
primarily contribute to the so-called ``reactive'' force. The
axial component of this reactive force was termed thrust. He
stated that the tangential component $u$ will have only small
contribution to the added mass and thus it will contribute
predominantly to the viscous resistance. The axial component of
the viscous force was termed drag. In this way, Lighthill
formulated the decomposition of the force on the fin into drag--
and thrust--causing mechanisms in spite of nonlinearities at high
Reynolds numbers. Note that the decomposition of the fin velocity into
$u$ and $w$ does not satisfy the realizability condition D2.
Lighthill developed expressions to obtain the reactive thrust due
to $w$ for elongated bodies or fins \cite{Ligh71a,Ligh90a}. Since
thrust $T$ was assumed to be caused by $w$, according to equation
\ref{eqn:w}, it is a function of $U_{w}-U$ \cite{Ligh90a}. The
thrust is directed in the forward direction, as desired, when
$U_{w}-U
> 0$, whereas it is directed backward when $U_{w}-U
< 0$. According to Lighthill, drag must be a function of $U_{w}$
and $(U_{w}-U)$ since it is caused by $u$ (equation
\ref{eqn:u}). However, no expressions for drag were developed.

\begin{table}[b!]
\centering

\begin{tabular}{M{2cm}  M{4cm}   M{4cm}}  \hline \hline

            &  \bf{Lighthill} & \bf{Current Study} \\ \hline \hline
\bf{Drag}  & 
\begin{align*}
u = \frac{U_{w}}{\cos[\alpha]} - \\
(U_{w} - U) \cos[\alpha]
\end{align*} 
& $u_{s} = \frac{U_{w}}{\cos[\alpha]}$  \\
\bf{Thrust} & $w = (U_{w}-U) \sin[\alpha]$ & 
\begin{align*}
w_{f} = (U_{w}-U)\sin[\alpha]  \\
u_{f} = -(U_{w}-U)\cos[\alpha]
\end{align*} \\
\hline \hline 
\end{tabular}
\caption{Comparison of Lighthill's kinematic decomposition and the one proposed here.}
\label{tab:lightus}
\end{table}

As shown in Fig.~\ref{fig:kd} 
and Table~\ref{tab:lightus}, below, 
the decomposition proposed in this work is different from Lighthill's. 
The velocities $u_{s}$ and $u_{f}$ that are
tangential to the wave surface in our slithering and frozen fin
modes, respectively, are given by
\begin{equation}
u_{s} = U_{w}/\cos[\alpha], \label{eqn:us}
\end{equation}
\begin{equation}
u_{f} = -(U_{w}-U)\cos[\alpha]. \label{eqn:uf}
\end{equation}
The tangential velocity in the slithering mode according to
equation \ref{eqn:us} corresponds only to the first part of the
tangential velocity $u$ according to Lighthill's decomposition in
equation \ref{eqn:u}. The tangential velocity in our frozen fin
mode (equation \ref{eqn:uf}) corresponds to the second part of the
tangential velocity $u$ in equation \ref{eqn:u}. The normal
velocity in the slithering mode is zero whereas that in the frozen
fin mode $w_{f}$ is given by
\begin{equation}
w_{f} = (U_{w}-U)\sin[\alpha], \label{eqn:wf}
\end{equation}
This normal velocity is the same as $w$ in equation \ref{eqn:w}.
Thus, our results suggest that part of the tangential velocity $u$
in the decomposition of Lighthill \cite{Ligh71a} that depends on
$(U_{w}-U)$ is in fact coupled with the normal component to
account for the thrust force according to the frozen fin mode.

In our decomposition the drag force is a function of $U_{w}$
unlike the consequence of Lighthill's decomposition where the drag
force is expected to be a function of $U_{w}$ and $(U_{w}-U)$. The
thrust force in our case is a function of $(U_{w}-U)$ like
Lighthill but it has an additional contribution from the
tangential velocity along the fin surface.

\section{Optimal height of the ribbon fin: Analysis}
\label{sec:optH}
\subsection{Swimming velocity}
In Fig.~\ref{fig:UsPl}b we presented the swimming velocity
obtained from self-propulsion simulations. In this section we will
show that it can also be obtained based on a reduced order model.
Doing so will provide insights into the mechanisms underlying the
simulation data. 

\begin{figure}[t!]
\centerline{\includegraphics[width=.49\textwidth]{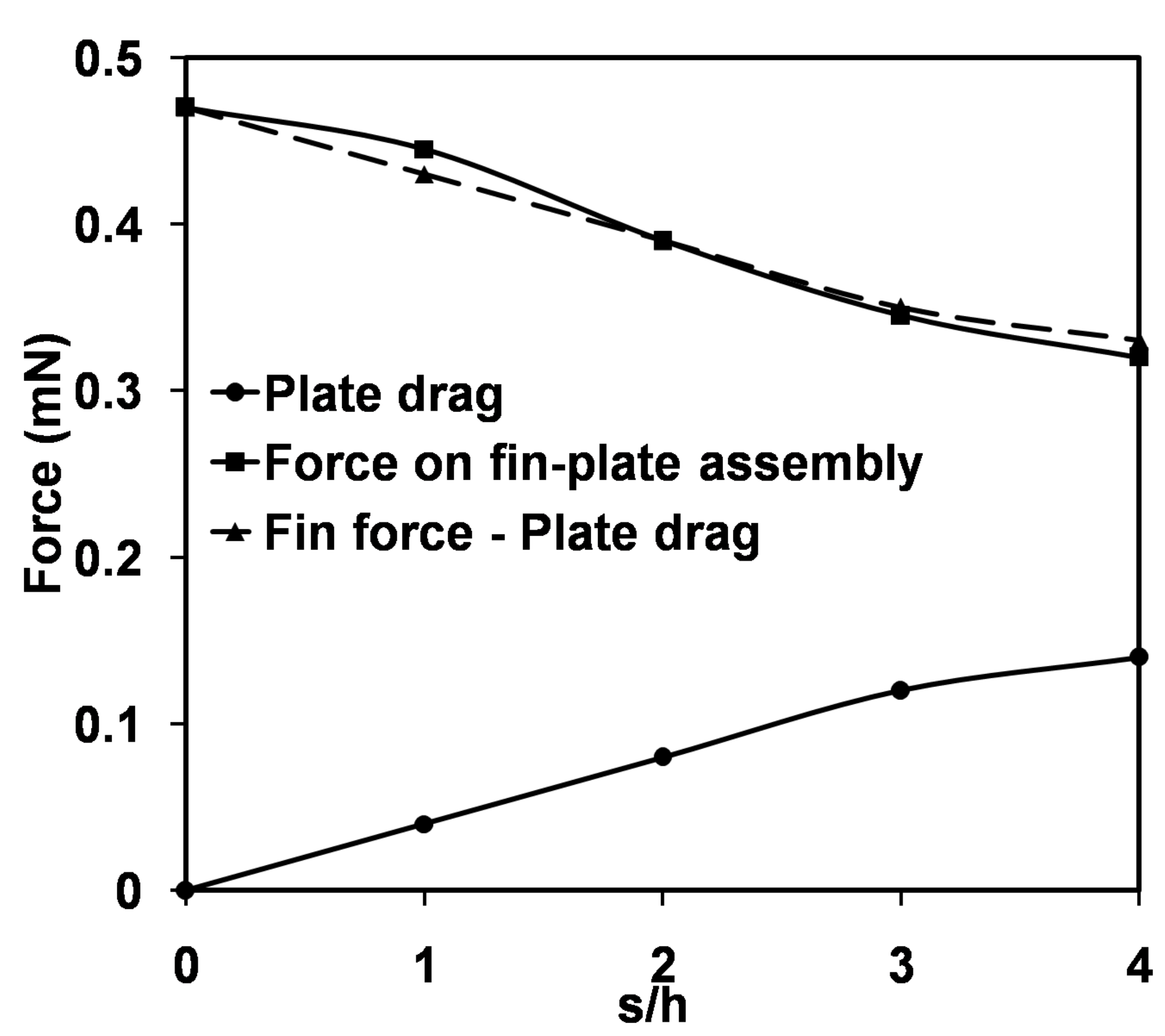}}
\caption{Force $F_{p}$ on the plate (---$\bullet$---), force
$F_{fp}$ on the fin-plate assembly (---$\blacksquare$---), and
$F_{f}-F_{p}$ (-- --$\blacktriangle$-- --) as a function of $s/h$.
Agreement between $F_{fp}$ and $F_{f}-F_{p}$ shows that there is
no momentum enhancement.} \label{fig:momenplot}
\end{figure}

The steady swimming velocity of a self-propelling
organism is the one at which the resultant force on the body,
averaged over a swimming cycle, is zero. In case of the fin-plate
assembly, forces on the plate and the fin are decoupled as seen in
Fig.~\ref{fig:momenplot}. Therefore, at steady swimming
$F_f[U_w,U,h]-F_p[U]=0$, where square brackets indicate
``function-of'' and $U$ is the translational velocity of the
plate-fin assembly. The value of $U$ that satisfies the force
equation is the swimming velocity $U_s$. Only $U_w, U$ and $h$ are the parameters of
interest in this analysis. Remaining parameters are
assumed to be constant.

According to the drag-thrust decomposition discussed in the Manuscript,
$F_{f}=T_{f}-D_{f}$, where $T_{f}$ and $D_{f}$ are the thrust and
drag forces on the ribbon fin, respectively. It was found that
$D_{f}$ depends on the wave velocity $U_{w}$ whereas $T_{f}$
depends on $(U_{w} - U)$. Using the drag-thrust decomposition of the
force $F_f$ on the fin, the force balance equation for the
plate-fin assembly during steady swimming can be written as
\begin{equation}
T_f[U_w-U,h]-D_f[U_w,h]-F_p[U]=0. \label{eqn:swim-gen}
\end{equation}
The functions for $T_f$, $D_f$, and $F_p$ are obtained as follows.
Fig.~\ref{fig:forces}c shows that $T_f$ depends on $(U_w-U)$. A
combination of linear and quadratic variations fit the graph well. 
Since the mechanism of thrust generation is
shown to be the low pressure caused by flow separation, thrust is proportional to the effective frontal
area of the flapping ribbon fin which scales as $h^2$. The drag
force $D_f$ is shown to be due to a boundary layer type flow caused by the
slithering motion (Fig.~\ref{fig:velpress}). It depends
on $U_w$ (Fig.~\ref{fig:forces}b), which is a constant in our simulations. The drag force
is proportional to the wetted surface area of the ribbon fin which
scales as $h$. Finally, $F_p$ is caused primarily by the boundary
layer on the plate. Hence, it is proportional to $U^{3/2}$ (all
other parameters are constant). Thus, equation \ref{eqn:swim-gen}
can be written in the following form
\begin{equation}
\{A_1(U_w-U)^2+A_2(U_w-U)\}h^2 - Bh - CU^{3/2}=0,
\label{eqn:swim-fun}
\end{equation}
where $A_1$, $A_2$, $B$, and $C$ are constants. The term involving
$h^2$ is the thrust force $T_f$, $D_f = Bh$,
and $F_p = CU^{3/2}$. Note that $U_w=15$~cm/s
for all cases considered here. Based on the data in Figures \ref{fig:forces}b and 
\ref{fig:forces}c, we find that $A_1 =
0.0059$~mN-s$^2$/cm$^4$, $A_2 = 0.0068$~mN-s/cm$^3$, and $B =
0.51$~mN/cm. We did simulations for flow over a plate and found
that $C = 0.017$~mN/(cm/s)$^{3/2}$. The only unknowns in equation
\ref{eqn:swim-fun} are $U$ and $h$. Thus, it can be used to obtain
the swimming velocity $U_s$ as a function of $h$. 


\textbf{\emph{Analytic solution for $U_s$:}} Now we solve the analytic equation
\ref{eqn:swim-fun} and compare it to the swimming velocity computed from 
self-propulsion simulations of the fin-plate assembly (Fig.~\ref{fig:UsPl}b).
It is possible to obtain an approximate closed form solution for
$h$ as a function of $U_s$ by solving equation \ref{eqn:swim-fun}.
To do so we note that equation \ref{eqn:swim-fun} is a quadratic
equation for $h$. Using the quadratic formula for $h$, keeping the
physically relevant solution, and using binomial expansion of a
square-root term up to first order, we get the following solution
\begin{equation}
h=\frac{B}{A_1(U_w-U_s)^2+A_2(U_w-U_s)}+\frac{CU_s^{3/2}}{B}.
\label{eqn:analy-swim}
\end{equation}
The first part on the right hand side corresponds to the swimming
of an isolated fin and the second part is the correction due to
the drag on the plate. 
The above equation is solved for $U_s$ and is compared to the values computed from
self-propulsion simulations in Fig.~\ref{fig:UsPl}b. Also plotted
in Fig.~\ref{fig:UsPl}b is the analytic solution for a
hypothetical freely swimming ribbon fin without a plate attached.
That solution is obtained from equation \ref{eqn:analy-swim} by
substituting $C=0$. Below a critical height $h_{c}$ the analytic
swimming velocity becomes negative. Analytic solution in this
regime is not plotted since the drag-thrust split model is not
valid in this regime. There is no self-propulsion
simulation data in this regime since the swimming velocity
is very small and the numerical accuracy is not sufficient. The
trends in Fig.~\ref{fig:UsPl}b will be discussed below in two
parts: \textbf{1)} trends for swimming velocity above the critical
fin height, and \textbf{2)} swimming at fin heights below the
critical value.

\textbf{1)} Fig.~\ref{fig:UsPl}b shows that, for fin heights
above a critical value, the swimming velocity from self-propulsion
simulations is in good agreement with the analytic curve for a
plate-fin assembly at smaller $h$ and agrees better with the
analytic curve for the fin-only case at larger $h$. This is
expected because at $h$ comparable to or larger than the plate
height of $s=2$~cm, the plate drag is dominated by the drag of the
ribbon fin. Hence, the plate-fin assembly swims at a velocity that
is close to the self-propulsion velocity of the ribbon fin itself.
The disagreement between simulated and analytic swimming
velocities of the plate-fin assembly, at larger $h$, arises
because the higher-order terms in the binomial expansion were
neglected in equation \ref{eqn:analy-swim}.

The key trend of $U_s$ is the rapid rise with respect to $h$ at
smaller $h$ and a slow change at larger $h$. 
This variation is the primary factor that determines the cost of transport (COT)
trends obtained in Fig.~\ref{fig:UsPl}c. It is clear from Figure
\ref{fig:UsPl}b that the trend in $U_s$ is inherent to the
self-propulsion velocity of the isolated ribbon fin itself. The
presence of the plate drag merely plays a role in shifting the
swimming velocity to a lower value. Equations \ref{eqn:swim-fun}
and \ref{eqn:analy-swim} imply that the root cause of the trend in
$U_s$ is the fact that the fin thrust $T_f$ scales as $h^2$ while
the fin drag $D_f$ scales as $h$, and the fact that the swimming
velocity affects only $T_f$. Due to this, at smaller $h$, any
increase in fin height increases the drag more than the thrust. To
achieve a balance between drag and thrust, the swimming velocity
must increase significantly so that the thrust is large enough to
equal the drag. For larger $h$ the effect is opposite. Any
increase in $h$ causes larger increase in thrust compared to drag.
Hence, to achieve drag-thrust balance the swimming velocity needs
to change only slightly.

\textbf{2)} The analytic solution leads to a critical height
$h_{c}$ at which the swimming velocity is zero (Figure
\ref{fig:UsPl}b). The analytic solution is not valid for this and
smaller heights because the drag-thrust split model for the fin
force is not accurate. The swimming velocities at fin heights
below the critical value, computed from self-propulsion
simulations, are very small indicating inefficient swimming, i.e.,
large COT. Thus, optimal swimming conditions do not fall within
this regime. We do not present these data because the numerical
accuracy was not sufficient to report them quantitatively. It was
computationally expensive for us to explore this regime because
very refined meshes are required in this case to resolve the flow
due to a tiny ribbon fin attached to a large plate.

\subsection{Cost of transport (COT)}

To undersand the COT trends we first consider the trend of average
power plotted in Fig.~\ref{fig:UsPl}a. It is seen that the data
fit a single curve that scales as $h^3$ over the entire range. It
is found from our computations that the power is dominated by the
work done to flap the ribbon fin laterally. The velocity $V_w$ of
lateral motion is maximum at the bottom edge of the ribbon fin and
scales as $h \theta_{max} f$. Thus, the power is expected to scale
as $\rho V_w^3 A$, where $A$ is some scale for area. The power
spent is not substantially influenced by the swimming velocity for
the scenarios considered here. Our computations also show that the
dominant contribution to power spent from lateral flapping comes
from the bottom edge of the fin. Thus, the appropriate scale for
$A$ is not expected to depend strongly on $h$. Therefore, it
follows that power $P \sim h^3$, consistent with Figure
\ref{fig:UsPl}a.

To compute the COT analytically we use the curve fit for power in
Fig.~\ref{fig:UsPl}a and divide it by the analytically obtained
swimming velocity (equation \ref{eqn:analy-swim}). This is plotted
in Fig.~\ref{fig:UsPl}c and compared with the COT computed from
self-propulsion simulations. The agreement is found to be good.

The analytic solution helps to better understand the trend in COT
$= P/U_{s}$ above the critical fin height. Since the swimming
velocity changes faster with respect to $h$ at lower values of
$h$, it compensates for the $h^{3}$ rise of power. This gives the
low and nearly constant trend of COT with respect to $h$ in the
simulation data. At larger $h$ the swimming velocity increases
with respect to $h$ at a rate that is less than $h^{3}$. Hence,
COT $= P/U_{s}$ increases with respect to $h$, thus, making
swimming with a larger fin less and less effective. As noted
before, the variation in the swimming velocity can be understood
in terms of different scalings, with respect to $h$, of the drag
and thrust of the ribbon fin. Therefore, the key factors that
explain the trends in COT with respect to the fin height are the
different mechanisms of drag and thrust on the ribbon fin.

\subsection{Effect of a realistic fish body drag}
In case of the idealized plate-fin swimmer, the plate drag is
dominated by boundary layer flow and scales as $U^{3/2}$. In case
of an actual fish, this may not be an accurate drag estimate due
to the presence of form drag on the body. To ensure that the key
features obtained in the idealized plate-fin model are not
affected by this factor, we consider an analytic solution with a
different body drag. To that end, we replace the plate drag term
$CU^{3/2}$ in equation \ref{eqn:swim-fun} with $EU^2$, which
represents the typical pressure drag scaling at high Reynolds
numbers. Our prior simulations for the drag force on a realistic
CAD model of the body of a black ghost knifefish \cite{MacI10a},
give $E=0.055$~mN-s$^2$/cm$^2$. Using this drag model for the body
we find the analytic solution for the swimming velocity. Then we
estimate the COT for the body-fin assembly by assuming that the
primary contribution to power comes from the fin -- which was
verified in the plate-fin case. To obtain COT, we use the same
power curve as that in Fig.~\ref{fig:UsPl}a  and divide it by the
analytically calculated swimming velocity based on the body drag
model. Fig.~\ref{fig:UsPl}c shows a comparison between the COT
for the plate-fin and body-fin assemblies. The key features
pertaining to the minimum in COT are the same. Thus, all the
trends discussed above about optimal fin height are applicable
even with a realistic body drag.

\section{Sensitivity of COT to perturbations in plate height}
\label{sec:pert}
We measured the body and the fin height of $13$ species from $8$ genera of South American 
weakly electric fishes in the family Apteronotidae 
(see Section \ref{sec:biodata} and Table \ref{tab:biodata}). 
We found that the mean body height, at mid-fin length, was $1.86$~cm with a standard deviation of $0.85$~cm. The mean 
fin height, at mid-fin length, was measured to be $0.97$~cm. This is within the range our prediction for 
optimal fin height. Although the standard deviation of the fin height is as high as $0.84$~cm, 
the measured fin height does not vary significantly. The standard deviation of fin 
height is $0.33$~cm. 

In this section we investigate how sensitive the COT trend is to 
variations in plate height. Note that the COT trend is the one that leads to the prediction of the fin height. For this, holding the fin height constant, 
we first find the rate of change of swimming speed with plate height using equation~\ref{eqn:analy-swim}.
The rate of change of velocity is then used to find the change in cost of transport due to perturbations in 
plate height. The change in cost of transport will provide insights into 
how optimal fin height will change with plate height. 

The rate of change of swimming velocity with plate height, after differentiating 
equation \ref{eqn:analy-swim} at constant fin height, is given by

\begin{eqnarray}
 &&{ \frac{\partial U_s}{\partial s} = -\frac{C^{\prime} U^{3/2} }{B} \left[ \frac{B(2A_1 (U_w-U_s)+A_2)}{( A_1 (U_w-U_s)^2 + A_2 (U_w-U_s) )^2} 
     +  \frac{3C\sqrt{U_s}}{2B} \right],    } 
     \label{eqn:duds}  
\end{eqnarray}
where $C^{\prime} = \partial C/ \partial s $. The drag on 
the plate has a linear dependence on the plate height. Consequently, $C$ will also have 
a linear dependence on plate height. We find $C^{\prime} = \partial C/ \partial s  = 0.0085$~mN-s$^{3/2}$/cm$^{3/2}$.  

The rate of change of cost of transport is found by taking the derivative of $P/U_s$ with respect to $s$ at constant fin height. It is given by
\begin{equation}
 \frac{\partial COT}{\partial s} = - \frac{COT}{U_s}\frac{\partial U_s}{\partial s}.
\end{equation}
The change in cost of transport, $\Delta COT$, due to a perturbation in plate height, $\Delta s$, is computed using 
the above equation. This is used to estimate the change in the computed value of the cost of transport corresponding to the standard 
deviation of the measured body height of 
fishes. These estimated changes are plotted in Figure~\ref{fig:UsPl}. It is seen that although the plate height does vary, 
the corresponding variation in COT is very small at small fin height $h$, while at larger values of $h$ the change is moderate. In short, 
the basic trend of the COT with fin height is unaltered by changes to the plate height.

\section{Navier-Stokes equation}

The Navier-Stokes equation for incompressible flow 

\begin{equation}
 \frac{\partial\bm{u}}{\partial t} + (\bm{u}\cdot\nabla)\bm{u} = -\frac{1}{\rho}\nabla p + \nu \nabla^2\bm{u},
 \label{eqn:NS}
\end{equation}
where $\bm{u}$ is the fluid velocity, $p$ is the pressure, $\nu$ is the kinematic viscosity of the fluid, and $\rho$ is the fluid density. The second term on the left-hand-side is the nonlinear inertia term referred to in the Manuscript.

\newpage



\end{document}